\documentclass{arxiv}
\usepackage[utf8]{inputenc}
\usepackage{graphicx}
\usepackage{float}
\usepackage{multirow}
\usepackage[colorlinks,allcolors=cyan!70!black]{hyperref}

\DeclareMathAlphabet{\mathcal}{OMS}{cmsy}{m}{n}

\title{Harnessing the Peripheral Surface Information Entropy from Globular Protein-Peptide Complexes}
\runningtitle{Harnessing Peripheral Surface Information Entropy} %% For page header

\author[1,2,*]{Tyler Grear}
\author[1,*]{Donald J. Jacobs}
\runningauthor{Grear and Jacobs} %% For page header

\affil[1]{Department of Physics and Optical Science, University of North Carolina at Charlotte, 9201 University City Boulevard, Charlotte, NC, 28223, USA}
\affil[2]{Department of Bioinformatics and Genomics, University of North Carolina at Charlotte, 9201 University City Boulevard, Charlotte, NC, 28223, USA}
\corrauthor[*]{tgrear@charlotte.edu and djacobs1@charlotte.edu}

\papertype{Article}
\begin{document}

%%%%%%%%%%%%%%%%%%%%%%%%%%%%%
%%%%%%%%%%%%%%%%%%%%%%%%%%%%%

\begin{frontmatter}

\begin{abstract}
Predicting favorable protein-peptide binding events remains a central challenge in biophysics, with continued uncertainty surrounding how nonlocal effects shape the global energy landscape. Here, we introduce \textit{peripheral surface information} (PSI) entropy, $S_{\Psi}$, a quantitative measure of the statistical variability in apolar and charged non-interacting surface (NIS) proportions across conformational ensembles. Using energy-directed molecular docking via HADDOCK3 and explicit-solvent molecular dynamics simulations, it is demonstrated that favorable binding partners exhibit emergent, low-entropy $\mathcal{N}$-states (discrete macrostates in NIS state space) indicative of preferential apolar/charged surface configurations. Across dozens of peptides and multiple receptor systems (WW, PDZ, and MDM2 domains), dominant $\mathcal{N}$-states persisted under varied docking parameters and initial conditions. An experimental meta-ensemble of WW domains from 36 high-resolution structures confirmed the presence of dominant NIS modes independent of \textit{in silico} methodology, suggesting an evolutionary selection pressure toward specific NIS fingerprints. These findings establish $S_{\Psi}$ as a thermoinformatic descriptor that encodes favorable binding constraints into unique statistical signatures of the NIS.
\end{abstract}

%%%%%%%%%%%%%%%%%%%%%%%%%%%%%

\begin{sigstatement}
The ability to reliably identify favorable protein-peptide binding events remains an unsolved problem in biophysics with implications for bioengineering, drug discovery, and our fundamental understanding of biomolecular recognition. We introduce \textit{peripheral surface information} (PSI) entropy, $S_{\Psi}$, a novel thermoinformatic measure that captures emergent organization of apolar and charged non-interacting surface regions across conformational ensembles. Uniting structural, energetic, and evolutionary perspectives, PSI entropy reveals conserved surface configurations which have been confirmed in 36 experimentally characterized complexes of the WW-domain model system, further supporting the observed phenomenon is independent of \textit{in silico} docking methodology. This work establishes a quantitative framework for exploiting peripheral surface organization in predictive modeling, enabling new strategies for active-site characterization and targeted peptide engineering.
\end{sigstatement}
\end{frontmatter}

%%%%%%%%%%%%%%%%%%%%%%%%%%%%%
%%%%%%%%%%%%%%%%%%%%%%%%%%%%%

\section*{Introduction}

\noindent Ideas that the non-interacting surface (NIS) during protein-protein association can contribute, in a non-negligible manner, to binding affinity calculations have progressed over the last few decades. They were largely debated within the context of allosteric effects and entropy-driven solvent reorganization \cite{kauzmann1959some,bacon1965nature}. In the mid-2000s, researchers began to discuss the possibility of an extended interface \cite{ladbury2004extended}, asking whether nonlocal interactions can meaningfully affect binding strength and/or specificity. Volumetric and calorimetric measurements, together with nuclear magnetic resonance (NMR) spectroscopy have enabled quantitative monitoring of these nonlocal influences on binding mechanisms. Additionally, changes in heat capacity upon formation of a complex can provide information on nonlocal effects \cite{ladbury2004extended,Jacobs2005Elucidating}. In systems governed by complex weak interactions, enthalpy-entropy compensation can limit the discriminative value of $\Delta G$-based descriptors \cite{cooper2001heat}. Free-energy functions that are overly concentrated on the binding interface often miss the global thermodynamic context provided by the environment. Here, the heterogeneity of free-energy contributions is related to information entropy, establishing that conformations corresponding to favorable energetics can be characterized by unique statistical signatures that are encoded on the peripheral surface.

Since the mid-2010s, researchers have called for quantitative descriptions of NIS characteristics in order to build a global model of binding affinity, seeking an Archimedean point for the problem of free-energy determination \cite{kastritis2013molecular,kastritis2013binding,visscher2015non}. Furthermore, new entropic considerations are expected for flexible complexes, where the correlation of buried surface area (BSA) and binding affinity does not hold \cite{kastritis2013binding}. This demands new global metrics derived from entire conformational ensembles rather than representative binding poses or energy profiles. It has been shown that the percentages of apolar and charged NIS (denoted $\mathcal{N}_{a}$ and $\mathcal{N}_{c}$ respectively) exhibit significant correlations with binding affinity \cite{kastritis2013molecular,kastritis2014proteins,vangone2015contacts,vangone2019large}, and NIS effects have been verified through alanine scanning \cite{kastritis2013molecular,kastritis2014proteins}. It has also been observed that proportions of $\mathcal{N}_{a}$ and $\mathcal{N}_{c}$ are conserved over orthologous complexes indicating an evolutionary selection pressure \cite{kastritis2014proteins,schweke2020protein}. Assuming evolution has performed the functional optimization \textit{a priori}, an effective peptide engineering protocol should harness the fingerprints that emerge from this selective pressure. A new information-theoretic entropy is proposed that encodes NIS properties over ensembles of molecular complexes consistent with favorable global binding conditions. This emergent peripheral surface information (PSI) entropy is also verified to exist independent of \textit{in silico} methods, extracted from ensembles of experimentally resolved molecular complexes.

%%%%%%%%%%%%%%%%%%%%%%%%%%%%%

\section*{Methods}

\noindent For each complex a per-residue solvent accessible surface area (SASA) was converted to relative solvent accessibility (RSA) using NACCESS residue maximum ASA values \cite{hubbard1993naccess}, see Supporting Material Table S1 for the full set of \(\mathrm{ASA}_{\max}\) values. Residues with \(\mathrm{RSA}\ge 0.05\) were assigned to three chemical classes: apolar (A), charged (C), and polar (P). Interface residues were identified geometrically by a 5.0~\AA\ heavy-atom distance cutoff between receptor Chain A and partner Chain B, and all such interacting residues were excluded. The remaining integer tuple of counts \((n_A,n_C,n_P)\) defines a macrostate label for each microstate, with NIS apolar and charged proportions denoted \(\mathcal{N}_{a}\) and \(\mathcal{N}_{c}\) respectively. Because the three chemical-class fractions are conserved (they sum to 1), the polar component was considered redundant once \(n_A\) and \(n_C\) were specified.

The docking software used to generate results was HADDOCK 3.0 (HADDOCK3) \cite{giulini2025haddock3}. This method was selected due to its modular and open-source framework allowing for customization of parameters/energy functions, providing maximal control over conformational ensemble generation. The standard HADDOCK3 pipeline consists of 3 steps: 1) rigid-body docking; 2) semi-flexible refinement in torsion angle space; and 3) molecular dynamics simulation (MDS) in explicit solvent. Parameters were initially set to the defaults unless explicitly stated in this description. Correspondent with the steps above, there are 3 energy-based HADDOCK scoring (HS) functions given by

\begin{equation}
HS = \begin{cases}
(0.01\cdot E_{vdw}) + (1.0\cdot E_{elec}) + (1.0\cdot E_{desolv}) + (0.01\cdot E_{AIR}) - (0.01\cdot BSA),& \text{if} \hspace{0.1mm} \text{ Step 1}\\
(1.0\cdot E_{vdw}) + (1.0\cdot E_{elec}) + (1.0\cdot E_{desolv}) + (0.1\cdot E_{AIR}) - (0.01\cdot BSA),& \text{if} \hspace{0.1mm} \text{ Step 2}\\
(1.0\cdot E_{vdw}) + (0.2\cdot E_{elec}) + (1.0\cdot E_{desolv}) + (0.1\cdot E_{AIR}),& \text{if} \hspace{0.1mm} \text{ Step 3}
\end{cases}
\label{eq8}
\end{equation}

\noindent where $E_{vdw}$ and $E_{elec}$ represent non-bonded van der Waals and Coulomb intermolecular energies respectively. $E_{desolv}$ is an empirical desolvation term, BSA is the buried surface area upon complexation in \AA$^{2}$, and $E_{AIR}$ is the restraint violation energy which expresses an agreement between experimental and back-calculated data. All energies were computed using the optimized potentials for liquid simulations (OPLS) force field \cite{jorgensen1988opls}, reported in units of kJ/mol.

\begin{figure}[h]
    \centering
    \begin{minipage}{0.39\textwidth} % Adjust width accordingly
        \includegraphics[width=1.75\textwidth]{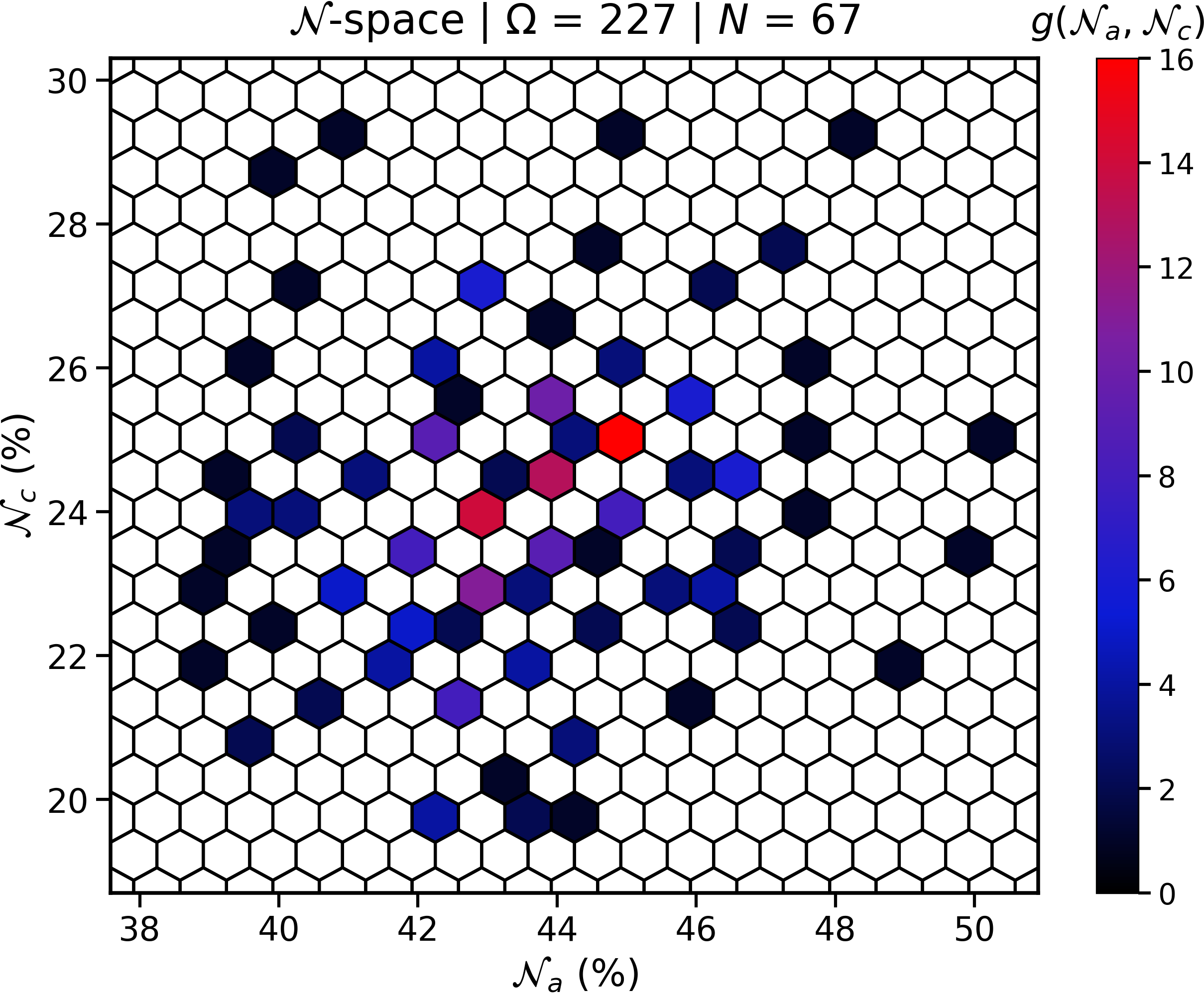}
    \end{minipage}%
    \hfill % Creates spacing
    \begin{minipage}{0.28\textwidth} % Space for caption
    %\vspace{0.0cm}
        \captionsetup{justification=justified} % Align left
        \caption{The $\mathcal{N}$-space hexbin plot generated under the \(\mathcal{N}\)-state (NIS macrostate) definition. Each hexagon represents a unit cell in the projected $(\mathcal{N}_{a},\mathcal{N}_{c})$ plane, while macrostates were defined by the exact integer tuple $(n_A,n_C,n_P)$ after RSA-thresholding and explicit interface exclusion. The exemplar ensemble contained $\Omega=227$ microstates that partitioned into $N=67$ distinct macrostates, with color indicating the multiplicity, $g(\mathcal{N}_{a},\mathcal{N}_{c})$. The $(\mathcal{N}_{a},\mathcal{N}_{c})$ coordinates were used for visualization, while the full $(n_A,n_C,n_P)$ tuple distinguishes macrostates. The existence of a non-uniform occupancy in $\mathcal{N}$-space indicates that, during an energy-directed search for bound conformations, a preferential subset of peripheral NIS patterns was populated.}
        \label{fig2}
    \end{minipage}
\end{figure}

For the purpose of establishing the methods, the exemplar model system was the WW domain (PDB ID 2LTW, experimental method NMR), a small highly flexible 36-residue globular protein characterized by two conserved tryptophan amino acids flanking each end. WW domains serve as a model system for understanding the thermodynamics and kinetics of the $\beta$-sheet fold \cite{pires2001solution}. The WW domain contains a binding groove with well-known peptide recognition \cite{lu2009knowledge}; namely, the Class I WW domain which recognizes PPxY or LPxY peptide motifs. The protein receptor was experimentally complexed with a 14-residue peptide: GESPPPPYSRYPMD. The first of 30 NMR conformers was selected, the protein and peptide were structurally separated then prepared and protonated using pdb2pqr \cite{jurrus2018improvements}. Next, independent structures were equilibrated in explicit TIP3P solvent using OpenMM \cite{eastman2023openmm} to decorrelate the structures from the experimentally bound conformations. The active residues were defined as (Tyr19, Leu21, His23, Gln26, and Trp30), that are discontinuous in sequence while being spatially contiguous, forming the known XP and Tyr grooves of the WW domain \cite{lu2009knowledge}. There were 3000 rigid-body conformations generated where the top-ranked 400 (according to Equation \ref{eq8}) were retained for the next stage. The second step of the docking protocol introduced flexibility to the interacting partners through a three-step MDS-based refinement in order to optimize interface packing. After the completion of Step 2, the top-ranked 400 models were further refined to remove all complexes having an interface root mean square deviation (iRMSD) greater than 10 \AA, resulting in 227 complexes in the exemplar ensemble. Dual energy-distribution plots are provided in Supporting Material Figure S1 that depict interfacial energetics over the generated ensembles described. The notable observation at this time was the self-organizing behavior of the NIS apolar and charged proportions ($\mathcal{N}_{a}$ and $\mathcal{N}_{c}$) as shown in Figure \ref{fig2}. The $(\mathcal{N}_{a},\mathcal{N}_{c})$ points span a two-dimensional space which contains 227 microstates ($\Omega=227$) that assembled into $N=67$ macrostates. The number of microstates that exist in a given macrostate is quantified by the multiplicity function, $g(\mathcal{N}_{a},\mathcal{N}_{c})$. This structured occupancy motivates an entropy-based description of NIS state variability.

Introduced by Claude Shannon in 1948, information entropy quantifies the amount of information contained in a message or the unpredictability (variability) of a system \cite{shannon1948mathematical}. For a system with a finite, countable number of microstates, $\Omega$, Shannon's information entropy is written as $S = - K \sum_{i} p_{i}\log_{2}(p_{i})$, where $K$ is a positive scaling (normalization) factor. Choosing $K=1$ gives the Shannon entropy in binary digits (bits). In this work, the unnormalized macrostate entropy is first defined in bits, followed by the introduction of a physically motivated $K$ from contact statistics. Each docking ensemble populates a discrete set of $\mathcal{N}$-states (macrostates) in \(\mathcal{N}\)-space. Let $\mathcal{N}_{i}$ be the $i$-th macrostate in \(\mathcal{N}\)-space with coordinates $(\mathcal{N}_{a},\mathcal{N}_{c})$, and take $p_{i}$ as the empirical probability $g(\mathcal{N}_{i})/\Omega$, where $g(\mathcal{N}_{i})$ counts how many microstates belong to that macrostate. This mapping of binding events to macrostate-level \(\mathcal{N}\)-space allows for a peripheral surface information (PSI) entropy, $S_{\Psi}$, to be expressed as

\begin{equation}
    S_{\Psi}^{'} = - K \sum_{i=1}^{N} \bigg ( \frac{g(\mathcal{N}_{i})}{\Omega} \bigg ) \log_{2} \bigg ( \frac{g(\mathcal{N}_{i})}{\Omega} \bigg )
\label{eq10}
\end{equation}

\noindent Where the prime notation indicates the intermediate unnormalized expression. $S_{\Psi}^{'}$ serves as a descriptor of the ensemble's statistical state rather than a direct component of the classical Gibbs free energy. The guiding hypothesis is that energetically favorable recognition is accompanied by reduced variability of peripheral NIS macrostates, and therefore a lower measured PSI entropy. This was treated as a working assumption rather than an \textit{a priori} guarantee, the practical question was whether \(S_{\Psi}\) separated favorable from unfavorable binding under controlled comparisons. Accordingly, this assumption was tested explicitly in the second Results Section via a cross-fertilization procedure, where biologically cognate (proper) and non-cognate (improper) peptides were docked to the same receptor under matched protocols.

%%%%%%%%%%%%%%%%

\subsection*{Normalization of $S_{\Psi}$}

Intermolecular contact statistics over complete ensembles were utilized to construct the normalization constant, $K$, for the expression in Eq.~\ref{eq10}. Each docking run yields an inter-chain contact table listing residue-residue pairs $(i,j)$ with an associated probability-like value treated here as an unnormalized "mass" of contact, $m_{i,j}$, defined from a C$\alpha$-C$\alpha$ contact probability. To incorporate the biophysical affinity of different contact types, a contact class label was assigned to each amino acid (e.g., apolar, polar, charged-positive, charged-negative). For a contact between class $(c_i)$ and $(c_j)$, a dimensionless scaling factor $\gamma(c_i,c_j)$ was applied yielding the contact class-weighted mass

\begin{equation}
  \tilde{m}_{i,j} = \gamma(c_i,c_j) \, m_{i,j}
  \label{eq3b}
\end{equation}

\noindent The mapping in Equation \ref{eq3b} generally upweights contacts known to be favorable (e.g., apolar-apolar, polar-polar, and opposite charges) and downweights unfavorable combinations (e.g., apolar-polar, apolar-charged, and like charges) with representative values $\gamma(c_i,c_j)\in[0.61,1.65]$, heuristically established from atomic contact statistics \cite{miyazawa1996residue}. The enthalpy-aware contact-classes, weights, and brief descriptions are exhaustively provided in Supporting Material Table S2.

Given an ensemble with nonzero contact masses, duplicate $(i,j)$ entries were summed over all members of the run to obtain a single class-weighted mass per distinct contact pair. Aggregating over the resulting set of distinct pairs defines an ensemble-level weighted total contact mass, $M = \sum_{(i,j)}\tilde{m}_{i,j}$. Taking the cardinality of the set of distinct $(i,j)$ pairs as $Q$, the following reciprocal relationship is defined as

\begin{equation}
  \frac{1}{K} = \frac{M}{Q}
  \label{eq_norm_constant}
\end{equation}

\noindent $M$ grows with consistent complementary interactions that concentrate on favorable contact modes shaped by hydrophobic packing and electrostatics. $M/Q$ is sensitive to recurrent binding motifs, while a large $Q$ with large $M$ signals energetic concentrations that are diffuse along the receptor surface. Inserting Equation \ref{eq_norm_constant} into Equation \ref{eq10} provides the explicit form

\begin{equation}
    S_{\Psi} = - \bigg ( \frac{Q}{M} \bigg ) \sum_{i=1}^{N} \bigg ( \frac{g(\mathcal{N}_{i})}{\Omega} \bigg ) \log_{2} \bigg ( \frac{g(\mathcal{N}_{i})}{\Omega} \bigg )
\label{eq_final_S_Psi}
\end{equation}

\noindent Each macrostate, $\mathcal{N}_i$, contributes $(-K\,p_i \log_{2} p_i)$ to $S_{\Psi}$. Subsequently, the relevant contributions are those from the \(\mathcal{N}\)-states rather than conventionally with microstate contributions. This was due to microstates only being considered through their multiplicities, $g(\mathcal{N}_i)$, and being otherwise indistinguishable, a strategic coarse graining of the information content. The unnormalized PSI entropy, $S_{\Psi}^{'}$, monitors how many distinct peripheral NIS patterns are explored by favorable binding poses in \(\mathcal{N}\)-space. Incorporating the normalization factor $K = Q/M$, where $M$ is the class-weighted total contact mass and $Q$ is the number of distinct contact pairs, results in an entropy per unit favorable contact enthalpy. The following regimes are described to provide an intuitive understanding for $S_{\Psi}$ as a measure of the global thermodynamic landscape of binding events.

\vspace{-0.25cm}

\paragraph{Regime I: low $S_{\Psi}$, low $Q/M$.} The interface is dominated by a focused set of high-affinity contacts that recurr across the ensemble, yielding small $Q/M$, and the emergent NIS patterns concentrate into a subset of macrostates such that $S_{\Psi}$ is small.

\vspace{-0.25cm}

\paragraph{Regime II: high $S_{\Psi}$, high $Q/M$.} Here, the class-weighted contact mass is spread over many weak or sporadic residue-residue pairs such that $Q/M$ is large; additionally, the ensemble populates a broad family of distinct \(\mathcal{N}\)-states giving a large $S_{\Psi}$. This regime represents high degeneracy and polyspecific recognition, consistent with transient or hub-like interactions.

\vspace{-0.25cm}

\paragraph{Regime III: intermediate $S_{\Psi}$, low $Q/M$.} In this intermediate $S_{\Psi}$ regime, the interface is built from strong recurrent contacts so $Q/M$ remains small, yet $S_{\Psi}$ is only moderately reduced, indicating the peripheral NIS patterns retain appreciable variability across binding poses. This corresponds to a landscape that is well-focused at the interface with strong contacts yet is globally adaptable at the periphery, allowing multiple \(\mathcal{N}\)-states to remain thermodynamically probable.

%%%%%%%%%%%%%%%%%%%%%%%%%%%%%
% =========================
% RESULTS
\section*{Results}
% =========================

\subsection*{Emergence of $\mathcal{N}$-States Under Diverse Conditions and Conformations}

\begin{figure*}[h!]
    \centering
    \includegraphics[width=1.0\textwidth]{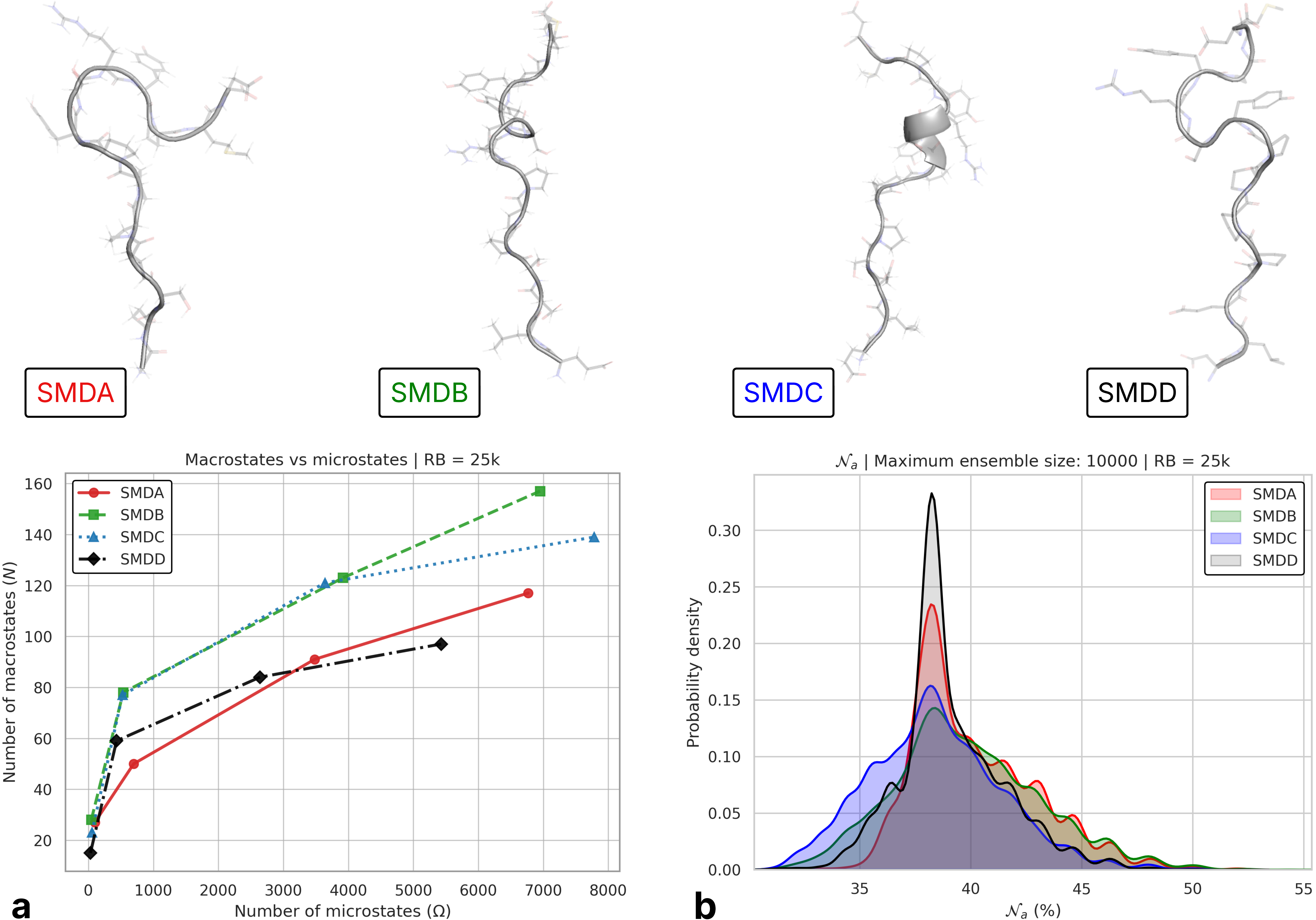}
    \caption{An overview of \(\mathcal{N}\)-state properties while varying: the Smad7 peptide conformations, number of rigid-body models, and number of models refined by semi-flexible MDS. (top row) Conformations are shown with N-termini on bottom and C-termini on top where backbone geometries are depicted in silver. (SMDA) Was taken from the NMR ensemble 2LTW without modification. (SMDB) This conformation was predicted with AF2. (SMDC) The prior AF2 conformation was equilibrated with MDS where a partial helix emerges. (SMDD) The final conformation was generated using AF3. (a) The number of macrostates, N, vs the number of microstates ($\Omega$) where the number of rigid-body (RB) models was 25000. As the number of microstates increases, the number of emergent macrostates follow a logarithmic trend, reaching a maximum of $N=157$ for SMDB. (b) Kernel density estimations of $\mathcal{N}_{a}$ proportions for the four Smad7 peptides. A dominant $\mathcal{N}_{a}$ mode emerges for all 4 peptide candidates, attaining a higher amplitude for the SMDA and SMDD partners.}
    \label{fig3}
\end{figure*}

Four conformations of the Smad7 peptide (GESPPPPYSRYPMD) were generated as shown along the top row of Figure \ref{fig3}. The first conformation (SMDA) was taken from the NMR ensemble (PDB ID 2LTW; conformer 1/30). The second conformation (SMDB) was predicted using AlphaFold2 (AF2). The third conformation (SMDC) resulted from performing an equilibration for 500 ps in explicit solvent on SMDB. The fourth peptide conformation (SMDD) was predicted with AlphaFold3 (AF3). This subset acts as an initial probe into whether the emergent $\mathcal{N}$-states are sensitive to peptide conformation. For all 4 conformations, the same protocols to dock Smad7 on the WW receptor were executed with two parameter sweeps of: \textit{i}) the number of rigid body models to generate in Step 1 with increments of (10000 and 25000); and \textit{ii}) the number of rigid models retained ($n_{ret}$) for Step 2 semi-flexible refinement were varied in maximal increments of (200, 1000, 5000, 10000).

The primary finding was that the organization into a subset of macrostates continues under the parameter sweeps. When the number of rigid-body models in Step 1 was 10000, the mean number of \(\mathcal{N}\)-states across all Smad7 peptide conformations was: $24.25 \pm{5.72}$ when $n_{ret}=200$, $65.50 \pm{12.60}$ if $n_{ret}=1000$, $106.75 \pm{16.68}$ at $n_{ret}=5000$, and $145.00 \pm{19.17}$ given $n_{ret}=10000$. This trend persisted when the number of rigid-body models was 25000. The smaller mean number of \(\mathcal{N}\)-states shows that the self-organization into subsets of macrostates occurs across the varied conformations, parameters, and conditions as exemplified in Figure \ref{fig3}\textcolor{cyan!70!black}{a}. Figure \ref{fig3}\textcolor{cyan!70!black}{b} shows the emergence of a dominant mode for each Smad7 ensemble. For a complete set of \(\mathcal{N}_{a}\) and \(\mathcal{N}_{c}\) distributions see Supporting Material Figures S2 and S3.

Next, the distributions of \(\mathcal{N}\)-states for a diverse set of WW domain-peptide complexes were evaluated to determine if common macrostates emerge. Step 3 from the standard HADDOCK pipeline is now included, executing MDS with explicit TIP3P solvent. The general expectation from the inclusion of Step 3 is that as the system relaxes there is an expected general increase in $S_{\Psi}$ across all systems due to thermal fluctuations during equilibration. This is expected to occur along with an increase in the number of \(\mathcal{N}\)-states due to an increased sampling of binding conformations, reported in Tables \ref{table1} and \ref{table2}. Two sets of peptide candidates were created, each used to generate ensembles of complexes with the WW domain. The first set, $\{PY\}$, contains 16 proline-rich peptides with motifs PPxY (9), LPxY (3), and PPGW (4). The second set, $\{RD\}$, is comprised of 16 randomly generated sequences where peptide lengths varied between 10 and 15. Each structure of $\{RD\}$ was predicted using AF3 then both sets were decorrelated from initial states by equilibrating with short MDS prior to initiating docking protocols. Full sequences for the $\{PY\}$ and $\{RD\}$ sets are provided in Supporting Material Tables S3 and S4 respectively. The number of Step 1 rigid-body models was 10000, and the number of retained models ($n_{ret}$) between steps was held fixed at $n_{ret}=200$.

\begin{table}[h]
\caption{$\mathcal{N}$-state properties between Step 2 (semi-flexible refinement) and Step 3 (MDS in explicit solvent) extracted from ensembles of protein-peptide complexes. Sixteen proline-rich peptides with motifs PPxY, LPxY, and PPGW were complexed with the WW domain (PDB ID 2LTW). The terms $\Delta \mathcal{N}$, $\Delta K$, and $\Delta S_{\Psi}$ indicate the Step 3 minus Step 2 changes in the number of \(\mathcal{N}\)-states, the normalization constant $K=Q/M$, and PSI entropy, respectively.}
\label{table1}
\setlength{\tabcolsep}{2pt}
\resizebox{1.0\textwidth}{!}
{
\begin{tabular}{|c|ccccccccc|ccc|cccc|}
\hline
 &  &  &  &  & PPxY &  &  &  &  &  & LPxY &  &  & \multicolumn{2}{c}{PPGW} &  \\ \cline{2-17} 
 & \multicolumn{1}{c|}{\small 2DJY} & \multicolumn{1}{c|}{\small 2KXQ} & \multicolumn{1}{c|}{\small 2LB1} & \multicolumn{1}{c|}{\small 2LB2} & \multicolumn{1}{c|}{\small 2LTV} & \multicolumn{1}{c|}{\small 2LTY} & \multicolumn{1}{c|}{\small 2LAW} & \multicolumn{1}{c|}{\small SMDA} & \small SMDC & \multicolumn{1}{c|}{\small 2EZ5} & \multicolumn{1}{c|}{\small 7LP5a} & \small 2MPT & \multicolumn{1}{c|}{\small 1ZCN} & \multicolumn{1}{c|}{\small 2ZQS} & \multicolumn{1}{c|}{\small 6JIZ} & \small 7BQF \\ \hline
\small $\Delta \mathcal{N}$ & \multicolumn{1}{c|}{8.00} & \multicolumn{1}{c|}{4.00} & \multicolumn{1}{c|}{3.00} & \multicolumn{1}{c|}{6.00} & \multicolumn{1}{c|}{4.00} & \multicolumn{1}{c|}{14.00} & \multicolumn{1}{c|}{12.00} & \multicolumn{1}{c|}{6.00} & 8.00 & \multicolumn{1}{c|}{2.00} & \multicolumn{1}{c|}{-13.00} & 8.00 & \multicolumn{1}{c|}{-3.00} & \multicolumn{1}{c|}{5.00} & \multicolumn{1}{c|}{-7.00} & -6.00 \\ \hline
\small $\Delta S_{\Psi}$ & \multicolumn{1}{c|}{0.22} & \multicolumn{1}{c|}{-0.84} & \multicolumn{1}{c|}{0.08} & \multicolumn{1}{c|}{-0.29} & \multicolumn{1}{c|}{0.22} & \multicolumn{1}{c|}{0.20} & \multicolumn{1}{c|}{0.06} & \multicolumn{1}{c|}{0.16} & 0.22 & \multicolumn{1}{c|}{-0.97} & \multicolumn{1}{c|}{0.02} & -0.48 & \multicolumn{1}{c|}{0.51} & \multicolumn{1}{c|}{-0.55} & \multicolumn{1}{c|}{-0.30} & -1.63 \\ \hline
\small $\Delta K$ & \multicolumn{1}{c|}{0.02} & \multicolumn{1}{c|}{-0.16} & \multicolumn{1}{c|}{0.01} & \multicolumn{1}{c|}{-0.06} & \multicolumn{1}{c|}{0.01} & \multicolumn{1}{c|}{-0.01} & \multicolumn{1}{c|}{0.00} & \multicolumn{1}{c|}{0.03} & 0.02 & \multicolumn{1}{c|}{-0.15} & \multicolumn{1}{c|}{0.02} & -0.10 & \multicolumn{1}{c|}{0.08} & \multicolumn{1}{c|}{-0.15} & \multicolumn{1}{c|}{-0.04} & -0.33 \\ \hline
\end{tabular}
}
\end{table}

\begin{table}[h]
\caption{$\mathcal{N}$-state properties between Step 2 (semi-flexible refinement) and Step 3 (MDS in explicit solvent) extracted from ensembles of protein-peptide complexes. Sixteen random sequences were predicted using AF3 then equilibrated with MDS in explicit TIP3P solvent (RD01 to RD16). The terms $\Delta \mathcal{N}$, $\Delta K$, and $\Delta S_{\Psi}$ indicate the Step 3 minus Step 2 changes in the number of \(\mathcal{N}\)-states, the normalization constant $K=Q/M$, and PSI entropy, respectively.}
\label{table2}
\setlength{\tabcolsep}{2pt}
\resizebox{1.0\textwidth}{!}
{
\begin{tabular}{|c|c|c|c|c|c|c|c|c|c|c|c|c|c|c|c|c|}
\hline
 & \small RD01 & \small RD02 & \small RD03 & \small RD04 & \small RD05 & \small RD06 & \small RD07 & \small RD08 & \small RD09 & \small RD10 & \small RD11 & \small RD12 & \small RD13 & \small RD14 & \small RD15 & \small RD16 \\ \hline
\small $\Delta \mathcal{N}$ & 15.00 & 8.00 & 6.00 & 9.00 & 9.00 & 13.00 & -6.00 & 1.00 & 4.00 & 1.00 & 14.00 & 2.00 & -1.00 & 14.00 & 4.00 & 7.00 \\ \hline
\small $\Delta S_{\Psi}$ & -1.38 & 0.23 & -1.13 & -0.85 & 0.21 & -0.96 & 0.23 & 0.10 & -0.18 & -0.15 & -0.45 & -1.65 & -0.62 & 1.11 & -0.01 & 0.34 \\ \hline
\small $\Delta K$ & -0.44 & 0.01 & -0.24 & -0.17 & 0.02 & -0.32 & 0.04 & 0.02 & -0.06 & -0.03 & -0.22 & -0.29 & -0.09 & 0.11 & -0.02 & 0.04 \\ \hline
\end{tabular}
}
\end{table}

Table \ref{table1} and Table \ref{table2} summarize how the number of \(\mathcal{N}\)-states, the normalized PSI entropy, and the normalization constant $K=Q/M$ evolve across systems. The number of \(\mathcal{N}\)-states more frequently increased than decreased from Step 2 to Step 3, consistent with added sampling during solvent relaxation. However, this trend was not universal as the condition $\Delta \mathcal{N}\geq0$ held for 26/32 ensembles, with decreases observed for 1ZCN, 6JIZ, 7BQF, 7LP5a, RD07, and RD13. The $S_{\Psi}$ levels at Step 2 and Step 3 separate the two peptide sets. At Step 2, $S_{\Psi}=2.95 \pm 1.09$ for $\{PY\}$ compared to $3.94 \pm 1.24$ for $\{RD\}$. At Step 3, $S_{\Psi}=2.74 \pm 0.98$ for $\{PY\}$ compared to $3.62 \pm 1.10$ for $\{RD\}$. In contrast, the unnormalized Shannon term ($S_{\Psi}^{'}$ with $K=1$) does not separate $\{PY\}$ vs $\{RD\}$. This indicates that the PY-RD separation is introduced by the $K=Q/M$ scaling rather than by macrostate diversity alone. For full tables and metrics, see Supporting Material Tables S5 and S6.

%%%%%%%%%%%%%%%%%%%%%%%%%%%

\subsection*{Robustness of PSI Entropy Across Varied Systems}

Two new receptors were included to monitor the \(\mathcal{N}\)-space trends already established with the WW domain. The first receptor was the mouse double minute 2 (MDM2) protein, a vital regulator of the tumor suppressing protein p53. The MDM2 structure (PDB ID 4HFZ) was resolved by X-ray diffraction (resolution 2.69 \AA) and originally complexed with an N-terminal p53 segment (ETFSDLWKLLP). The second receptor was the postsynaptic density-95/discs large/zonula occludens-1 (PDZ) protein (PDB ID 1ZUB) attributed with the organization of neurotransmitters and signaling pathways at neuronal synapses. The structure of 1ZUB was resolved by NMR where the first of 20 conformers was selected. The 1ZUB protein was complexed with an ELKS1b C-terminal peptide, CDQDEEEGIWA. The receptor and peptide of 4HFZ/1ZUB were separated then decorrelated from their initial conformations by: 1) equilibrating the receptors in explicit solvent using OpenMM; and 2) peptide conformations were diversified by performing a 20 ns simulation on each in TIP3P solvent then sampling every 2 ns resulting in 10 conformations for each peptide candidate. The active sites provided to HADDOCK3 were expanded compared to the previous section with the WW domain (5 active residues). For MDM2 the active site was set to 54-99 (45 residues), containing the known binding groove that is lined with apolar amino acids. When peptides were docked to the PDZ protein, active sites were defined as 623-639 and 681-695 (a total of 32 residues) covering the binding groove formed by a carboxylate binding loop, $\beta B$ beta strand, and $\alpha B$ helix. A cross-fertilization procedure was performed to test the working assumption established in the Methods following Eq.~\ref{eq10} (see Supporting Material Table S8 for a summary of pairings). This can be succinctly stated as: given a biologically cognate (proper) and non-cognate (improper) peptide candidate docked to the same protein receptor, a favorable candidate should measure a lower $S_\Psi$ compared to candidates that do not bind well. The procedure involved 2 phases where each phase utilized a different protein receptor, a total of 40 HADDOCK3 runs (10 proper and 10 improper runs per phase). The ambiguous interaction restraints (AIRs) were removed from Step 3 during MDS. This acts as a molecular stress test for favorable interactions; consequently, a decrease in $\Omega$ of final ensembles was expected.

Proper and improper ensembles are summarized separately at the phase level. The unnormalized Shannon term ($S_{\Psi}^{'}$ with $K=1$) is computed per run from the macrostate probability distribution \(\{p_i\}\), then averaged across 10 replicas. The scaling coefficient, $K$, is computed from contact statistics by pooling the nonzero contact pairs (edges) and their contact weights over all replicas in the phase. The reported phase-level PSI entropy is then \(S_{\Psi}^{\mathrm{phase}} = K_{\mathrm{phase}}\langle S_{\Psi}^{'}\rangle_{\mathrm{runs}}\), ensuring that the normalization reflects the collective binding-site contact structure of the phase rather than an average of per-run normalizers. Pooling is appropriate because both $Q$ and $M$ depend on the support and weights of the contact network, which are estimated more stably at the phase level than from any single replica. Averaging per-run ratios would instead conflate run-to-run sampling variability in contact support with the phase-level normalization itself.

\begin{figure*}[h!]
    \centering
    \includegraphics[width=1.0\textwidth]{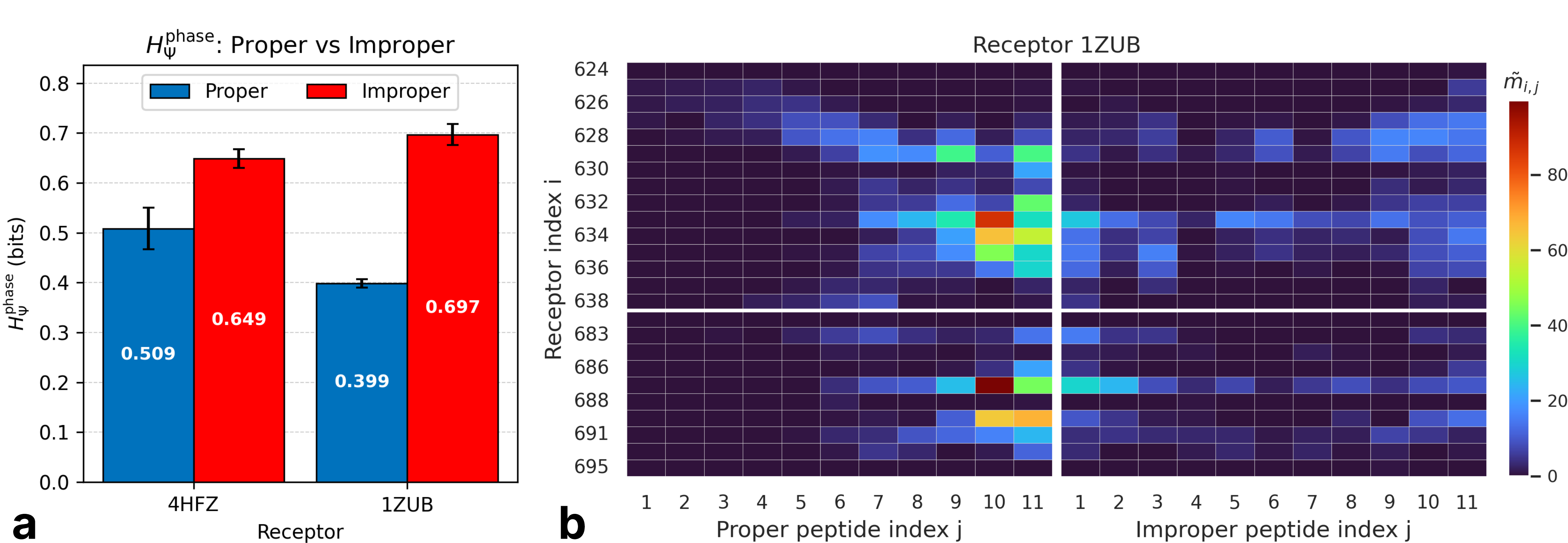}
    \caption{An overview for the cross-fertilization procedure. (a) Under matched protocols, proper ensembles exhibit lower $S_{\Psi}$ than improper ensembles for both receptors. Note, the separation \(S_{\Psi}^{\mathrm{phase}}(\mathrm{proper}) < S_{\Psi}^{\mathrm{phase}}(\mathrm{improper})\) persisted even when considering uncertainity (shown by error bars). (b) The enthalpy-weighted contact masses with an upweighting of spatially-dense favorable contacts for the proper ELKS1b peptide engagement. This exemplifies Regime 1 with low $S_{\Psi}$ and low $Q/M$.}
    \label{fig_Results2A}
\end{figure*}

Across 10 replicas per phase, ensemble sizes and macrostate counts varied by receptor and condition. Despite replica-to-replica measurement uncertainty, the ordering \(S_{\Psi}^{\mathrm{phase}}(\mathrm{proper}) < S_{\Psi}^{\mathrm{phase}}(\mathrm{improper})\) was preserved for both receptors. For 4HFZ, the proper ensembles had $\Omega = 47.6 \pm 32.0$ and $N = 42.2 \pm 26.4$, while the improper ensembles had $\Omega = 50.1 \pm 19.5$ and $N = 46.1 \pm 16.4$. For 1ZUB, the proper ensembles had $\Omega = 63.1 \pm 17.7$ and $N = 58.3 \pm 15.3$, while the improper ensembles had $\Omega = 24.7 \pm 8.8$ and $N = 23.2 \pm 7.2$. Despite this variability, the phase-level $S_{\Psi}$ values followed the working assumption under matched protocols, as shown in Figure \ref{fig_Results2A}\textcolor{cyan!70!black}{a}. For the 4HFZ receptor phase, $S_{\Psi}^{phase}=0.5086$ (proper) versus $S_{\Psi}^{phase} = 0.6489$ (improper), corresponding to $\Delta\%=21.6\%$. For the 1ZUB receptor phase, $S_{\Psi}^{phase}=0.3986$ (proper) versus $S_{\Psi}^{phase} = 0.6969$ (improper), with $\Delta\%=42.8\%$. Here, $\Delta\% = (1 - S_{\Psi}^{proper}/S_{\Psi}^{improper})\times 100$.

The enthalpy-weighted contact maps shown in Figure \ref{fig_Results2A}\textcolor{cyan!70!black}{b} refine this view by spatially localizing the interactions of proper ensembles. Supporting Material Figures S4 and S5 display cross-fertilization contact map probabilities, \(\rho(i,j)\), contact-class weighted probability maps, \(\tilde{m}_{i,j}\), and the ensemble-level net contact-class modulation, \(\sum_{\Omega} (\gamma(c_{i},c_{j})-1.0) \cdot m_{i,j}\), for 4HFZ and 1ZUB receptor phases. For the ELKS1b C-terminal peptide engagement, the contact class-weighted mass, \(\tilde{m}_{i,j}\), collapses to a compact patch of favorable contacts in the canonical groove, explicitly upweighted by \(\gamma(c_{i},c_{j})\). This spatial focusing exemplifies Regime 1 (low $S_{\Psi}$ and low $Q/M$), where a small subset of contacts carries a disproportionate share of the total engagement. The contact masses that dominate $\tilde{m}_{i,j}$ also dominate the contributions to $S_{\Psi}$; subsequently, Figure \ref{fig_Results2A} links reduced PSI entropy directly to a concentrated pattern of biophysically favorable contacts on the active site. The improper ensemble in Figure \ref{fig_Results2A}\textcolor{cyan!70!black}{b} falls into Regime 2 with elevated $Q/M$ and larger $S_{\Psi}$ per unit contact density. For cross-fertilization calculations of statistics and error, see Supporting Material Calculations S1 and S2 respectively.

Pooling all replicas within each phase, the phase-pooled \(\mathcal{N}\)-space densities provide a distribution-level view of peripheral organization under matched protocols (see Supporting Material Figures S6-S7). Because \(\Omega\) differs across conditions, both the absolute effective macrostate count (\(N_{\mathrm{eff}}=2^S\)) and its size-normalized fraction (\(N_{\mathrm{eff}}/\Omega\)) are used to distinguish changes in probability mass. For 4HFZ, the proper ensemble is more concentrated in \(\mathcal{N}\)-space, with a smaller effective number of macrostates (\(N_{\mathrm{eff}}=220.5\) for proper versus \(258.2\) with improper). For 1ZUB, the dominant \(\mathcal{N}\)-space mode is preserved while the normalized effective state fraction decreases (\(N_{\mathrm{eff}}/\Omega=0.481\) for proper versus \(0.629\) with improper). These pooled distributions show that reduced \(S_{\Psi}^{\mathrm{phase}}\) coincides with a tightening of peripheral state occupancy in a receptor-dependent manner. For the corresponding definitions of \(N_{\mathrm{eff}}\) and \(N_{\mathrm{eff}}/\Omega\) under phase pooling, see Supporting Material Calculations S3.

%%%%%%%%%%%%%%%%%%%%%%%%%%

%\clearpage
\subsection*{Experimental Validation of Dominant $\mathcal{N}$-States}

\begin{figure*}[h!]
    \centering
    \includegraphics[width=1.0\textwidth]{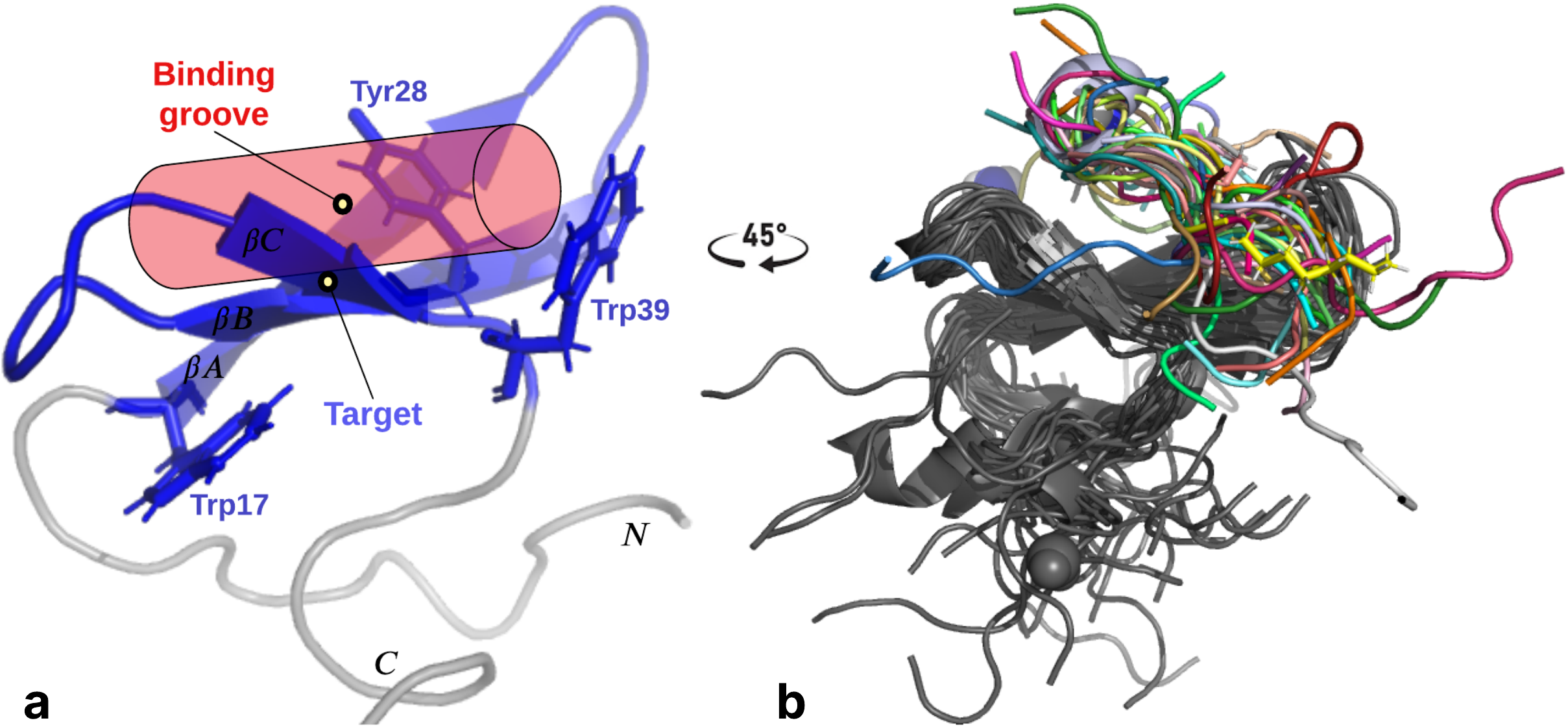}
    \caption{The WW domain side (a) and front (b) views. (a) The binding groove of the WW domain exists within the blue region bound by two highly conserved tryptophans (Trp17 and Trp39). Taking on the activated conformational state, the Tyr28 and Trp39 amino acids exhibit upward orientations relative to the beta sheet, likely due to $\pi$ effects which are known contributors of favorable binding enthalpy. (b) An alignment and superposition of all 36 PDB entries (34 NMR, 2 X-ray diffraction). The varying WW receptors (gray) are complexed with their corresponding peptide, ligand, or small molecule (multicolored ribbons).}
    \label{fig10}
\end{figure*}

To investigate whether the emergence of dominant \(\mathcal{N}\)-space modes is a general phenomenon independent of \textit{in silico} docking, ensembles of experimentally resolved structures were analyzed. The goal is to identify conserved surface configurations that may emerge independent of simulation protocols. Such convergence of NIS properties with specific interfaces would suggest an evolutionary selection pressure. This link between interfacial properties and selection pressure has been discussed previously \cite{kastritis2014proteins,schweke2020protein}, though not through the lens of NIS composition. For demonstration purposes, the WW domain is again used as a model system. The PDB RESTful API was utilized in Python to extract all structures with Pfam Protein Family (PF00397) annotations related to the WW domain. This query resulted in a total of 206 entries resolved from 4 experimental techniques: Solution NMR (107), X-ray Diffraction (96), Electron Microscopy (2), and Solid-State NMR (1). The 206 entries were spread over the following organisms: Homo sapiens (166), Mus musculus (25), Rattus norvegicus (4), Drosophila melanogaster (3), Saccharomyces cerevisiae (3), and others (5). Further refinement to all monomer WW domains complexed with small molecules, ligands, and peptides (including engineered molecules) resulted in 36 entries, 34 from NMR and 2 from X-ray experiments (see Supporting Material Table S7) that were then pooled into a meta-ensemble of \(\Omega=657\) microstates (complexes). The result of an alignment (reference structure 2LTW, Figure \ref{fig10}\textcolor{cyan!70!black}{a}) followed by a superposition is presented in Figure \ref{fig10}\textcolor{cyan!70!black}{b}.

\begin{figure*}[h!]
    \centering
    \includegraphics[width=1.0\textwidth]{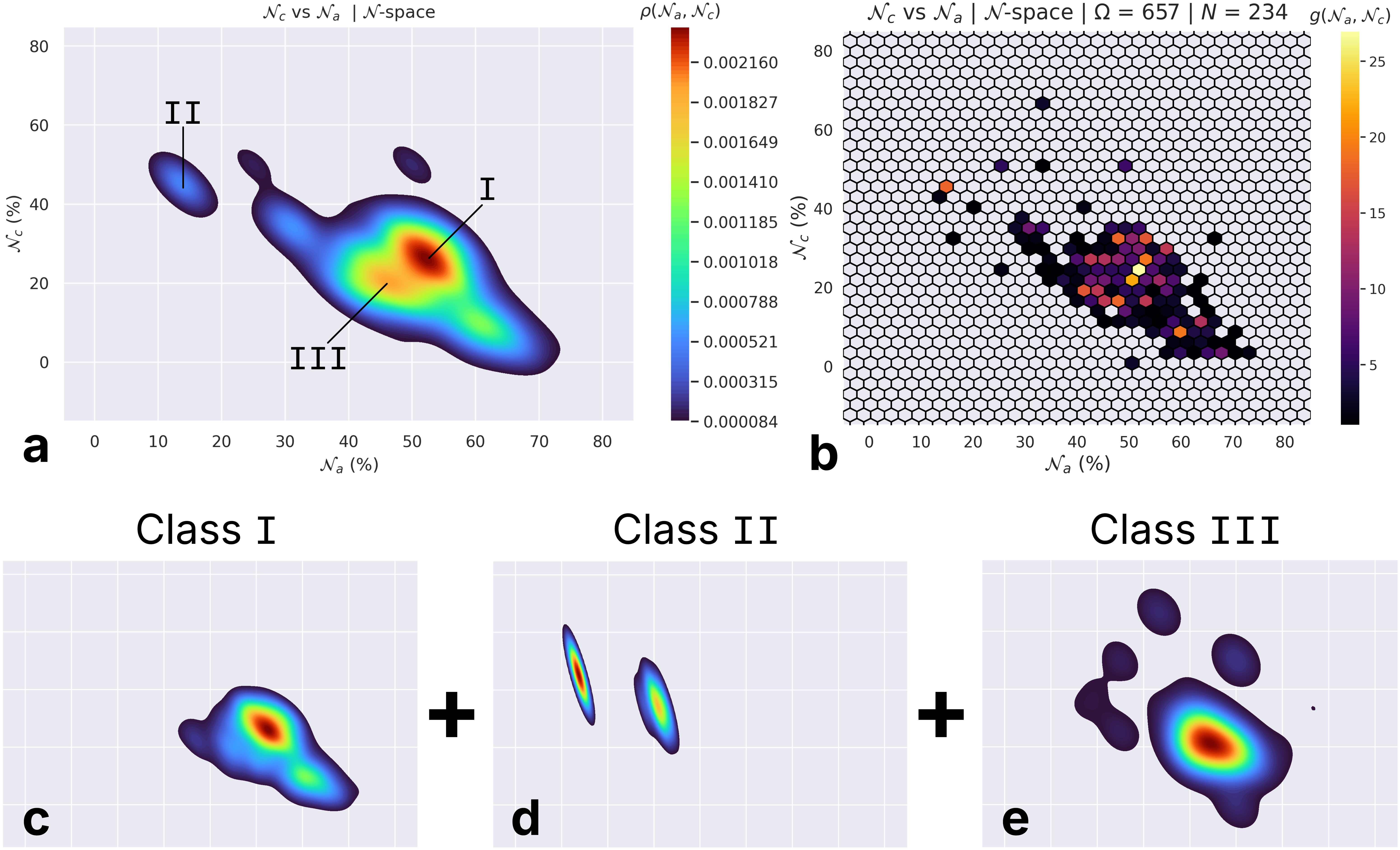}
    \caption{Bivariate distributions derived from a meta-ensemble of varying experimental WW domains complexed with small molecules, ligands, and peptides. (a) The continuous $\mathcal{N}$-space containing all 657 curated experimental protein-partner complexes. It is readily apparent that a dominant mode of the bivariate distribution exists, shown by red/dark-red regions. (b) The discrete $\mathcal{N}$-space (hexbin plot) explicitly shows 3 general regions of high multiplicity across all microstates. Trifurcating the meta-ensemble based on known classification, the corresponding bivariate distributions are shown in (c-e). The dominant class contributions to the meta-ensemble are indicated by (I, II, and III) in (a).}
    \label{fig11}
\end{figure*}

The multi-colored ribbons shown in Figure \ref{fig10}\textcolor{cyan!70!black}{b} exemplify a variety of partners in terms of both binding poses and amino acid composition. This conformational messiness was intentionally curated as the expected emergence of a dominant $\mathcal{N}$-state should occur along with statistical outliers and intrinsic conformational variance. Using the same process as described in the previous sections, the corresponding $\mathcal{N}$-space was constructed from the experimental meta-ensemble. A kernel density estimation was performed for the variables $\mathcal{N}_{a}$ and $\mathcal{N}_{c}$, spanning a two-dimensional continuous $\mathcal{N}$-space. The bivariate distribution of all 657 microstates is presented in Figure \ref{fig11}\textcolor{cyan!70!black}{a} where a dominant mode in $\mathcal{N}$-space is confirmed by a dense area of probability. The discrete \(\mathcal{N}\)-space is shown in Figure \ref{fig11}\textcolor{cyan!70!black}{b} where 3 regions of higher multiplicities are revealed. There are three well-known classes of WW domains based on peptide recognition \cite{lu2009knowledge}; accordingly, the meta-ensemble was trifurcated and the bivariate distributions were calculated for each class (Figures \ref{fig11}\textcolor{cyan!70!black}{c}-\textcolor{cyan!70!black}{e}), see Supporting Material Figure S8 for full plots with axis labels. A dominant mode was observed for both the WW family as a whole and each individual class. These insights preliminarily support the hypothesis that as globular protein-peptide systems tend towards specificity over evolutionary timescales, the NIS properties converge towards a dominant mode in $\mathcal{N}$-space, phenomena that are driven by evolutionary selection pressure.

The effective macrostate fraction, \(N_{\mathrm{eff}}/\Omega=0.246\) (\(N_{\mathrm{eff}}=161.8\)), is consistent with state occupancy being concentrated rather than spreading uniformly across the experimental space. Notably, \(N=234\) distinct \(\mathcal{N}\)-states (macrostates) were observed where 103 were singletons (\(g_i=1\)), while the maximum macrostate occupancy reached \(g_{\max}=20\). This heavy-tailed occupancy profile indicates that most states are sparsely sampled whereas a small subset repeatedly concentrates probability mass into dominant modes. Analysis of the \(g_{\max}=20\) macrostate indicates that its membership is not restricted to a single experimental ensemble, it is partitioned across multiple PDB entries (e.g., 2EZ5 and 2M3O); furthermore, 131 macrostates (${\sim}56\%$) contain microstates from more than one NMR ensemble, confirming that the preferred \(\mathcal{N}\)-space configurations are robust to varied experimental determinations over 15 years. Taken together, these results establish a consistent dominant-mode picture in \(\mathcal{N}\)-space ranging from energy-directed \textit{in silico} ensembles to experimentally resolved conformational diversity.

%%%%%%%%%%%%%%%%%%%%%%%%%%%%%%%%%%%%%%

\section*{Discussion}

Across docking protocol parameters, cross-fertilized receptors, and the experimental WW-domain meta-ensemble, peripheral surface organization repeatedly collapses toward preferred \(\mathcal{N}\)-state regions. This convergence indicates that NIS composition is constrained in a manner that is coupled to favorable recognition and possibly to evolutionary selection pressure \cite{kastritis2013binding,visscher2015non}. This work directly answers the longstanding call-to-action for quantitative metrics capable of encoding non-interacting surface characteristics into a global model of recognition \cite{kastritis2013molecular,kastritis2013binding,visscher2015non}. By providing a formal information-theoretic basis for measuring the statistical variability of nonlocal effects, \(S_{\Psi}\) explicitly realizes the abstraction of classical enthalpy-entropy compensation into an alternative perspective between interfacial energetics and information entropy. In this view, \(S_{\Psi}\) adopts the role of a molecular informational analog to Boltzmann’s \(H\)-theorem \cite{boltzmann1872weitere,shannon1948mathematical}.

The separation in normalized PSI entropy observed between biologically cognate and non-cognate complexes, both in the curated PY/RD sets and under stringent cross-fertilization protocols, demonstrates potential as a discriminatory metric. Analysis of the experimental WW-domain meta-ensemble further supports the collapse of peripheral organization into preferential NIS modes with the normalized effective macrostate fraction of the pooled meta-ensemble (\(N_{\mathrm{eff}}/\Omega \approx 0.25\)) being significantly lower than the median of individual NMR ensembles (\(N_{\mathrm{eff}}/\Omega \approx 0.36\)). This trend confirms that when viewed across the entire experimental landscape, \(\mathcal{N}\)-state occupancy remains highly concentrated rather than spreading uniformly, a result that exceeds the expectations of a simple additive model. Beyond serving as a diagnostic tool, the patterns revealed in \(\mathcal{N}\)-space constitute unique ensemble-level fingerprints that capture the statistical signature of a complex. Unlike conventional structural descriptors that focus on individual snapshots, these fingerprints represent a distribution-level property. Such biological signatures are well-suited for integration into next-generation machine learning models, where PSI entropy moves predictive modeling toward a more holistic description of structural, evolutionary, and thermodynamic constraints.

Looking forward, the ability to monitor trajectories in \(\mathcal{N}\)-space offers a powerful new lens for identifying biophysical patterns under varying environmental conditions, such as the MDM2 protein under osmotic shock. These trajectories provide explicit design targets for the rational modulation of protein-solvent interactions. For instance, rather than optimizing for high-affinity binding, an "anti-directed" protocol could target the destabilization of complexes or the disruption of allosteric pathways by driving \(\mathcal{N}\)-space occupancy away from dominant modes, thereby increasing \(S_{\Psi}\). This shifts the focus from simple pose-searching to an iterative mutational scanning framework where the thermoinformatic measure, \(S_{\Psi}\), serves as an objective function. Through the engineering of peripheral surface fingerprints, PSI entropy enables a design paradigm that treats the non-interacting surface as a controllable handle for tuning the stability and function of macromolecular interactions.

\section*{Author Contributions}

T.G. and D.J.J. designed the research. T.G. carried out all simulations, analyses, and writing. Both T.G. and D.J.J. contributed to the theoretical development of PSI entropy. Funding was acquired by D.J.J.

%%%%%%%%%%%%%%%%%%%%%%%%%%%%%%%%%%

\section*{Funding Statement}

Research reported in this publication was supported by the National Institute of General Medical Sciences of the National Institutes of Health under Award Number R15GM146200. The content is solely the responsibility of the authors and does not necessarily represent the official views of the National Institutes of Health.

%%%%%%%%%%%%%%%%%%%%%%%%%%%%%%%%%%

\section*{Acknowledgments}

We thank the UNC Charlotte University Research Computing (URC) group for facilitating the use of high-performance computing (HPC) clusters and systems.

%%%%%%%%%%%%%%%%%%%%%%%%%%%%%%%%%%

\section*{Declaration of Interests}

The authors declare no competing interests.

%%%%%%%%%%%%%%%%%%%%%%%%%%%%%%%%%%

\section*{Supporting Material}

Supporting Material is included at the end of this document.

% Uncomment if using bibtex (default)
\bibliography{bibliography}

%%%%%%%%%%%%%%%%%%%%%%%%%%%%%%%%%%
%%%%%%%%%%%%%%%%%%%%%%%%%%%%%%%%%%
\clearpage

% ======================================================================
% Supporting Material — use the original BJ-style frontmatter
% (matching `supporting.tex` title page formatting)
% ======================================================================

% ---------- S-numbering for Supporting Material ----------
\setcounter{figure}{0}
\setcounter{table}{0}
\setcounter{equation}{0}
\renewcommand{\thefigure}{S\arabic{figure}}
\renewcommand{\thetable}{S\arabic{table}}
\renewcommand{\theequation}{S\arabic{equation}}
\renewcommand\thesubsection{S\arabic{subsection}}

% ---------- Title & Author Info (verbatim from `supporting.tex`) ----------
\title{Supporting Material for: \\ \emph{Harnessing the Peripheral Surface Information Entropy from Globular Protein-Peptide Complexes}}
\runningtitle{Supporting Material}

% IMPORTANT: Do NOT re-declare \author/\affil here.
% The `authblk` package appends authors/affils, which would duplicate them on
% the Supporting Material title page. We reuse the already-declared main
% manuscript author/affiliation block, since it is identical.

% ---------- Front matter (verbatim from `supporting.tex`) ----------
\begin{frontmatter}
\begin{abstract}
This PDF contains Supporting Material for the above manuscript, including additional figures (S1-S8), tables (S1-S8), and calculations (S1-S3). Supporting items are cited in the main text (i.e., "Supporting Material Figure S1").
\end{abstract}
\end{frontmatter}

\section*{Contents}
\begin{center}
\begin{tabular}{@{}p{0.7\textwidth}r@{}}
\large Supporting Figures & \large pp.\ \pageref{supp:figures:start}--\pageref{supp:figures:end} \\
\large Supporting Tables & \large pp.\ \pageref{supp:tables:start}--\pageref{supp:tables:end} \\
\large Supporting Calculations & \large pp.\ \pageref{supp:calcs:start}--\pageref{supp:calcs:end} \\
\end{tabular}
\end{center}

\newpage

%%%%%%%%%%%%%%%%%%%%%%%%%%%%%%%%%%%%%%%%%%%%%
% ---------- Supporting Figures ----------
\section*{Supporting Figures}
\phantomsection
\label{supp:figures:start}

\begin{figure}[H]
    \centering
    \includegraphics[width=1.0\textwidth]{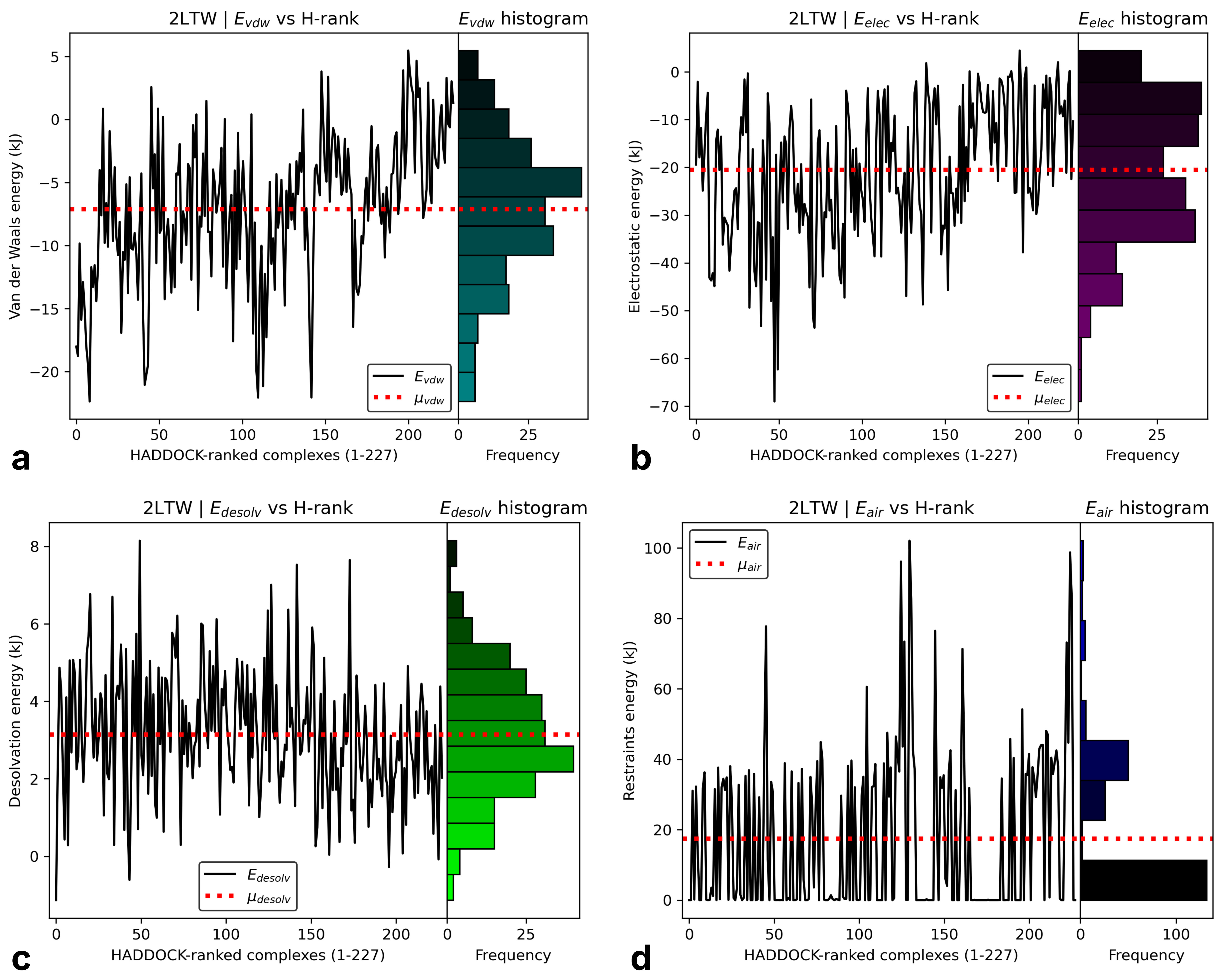}
    \caption{Dual energy-distribution plots for the WW domain-GESPPPPYSRYPMD conformational ensemble. The x-axes represent a 227 dimensional state space where each conformational state is ranked by HADDOCK score. The y-axes correspond to the energies: (a) van der Waals; (b) electrostatics; (c) desolvation; and (d) restraint violations. A Gaussian distribution indicates consistency of that energy over the ensemble of complexes as with (c).}
    \label{figS1}
\end{figure}

\newpage

%%%%%%%%%%%%%%%%%%

\begin{figure}[H]
    \centering
    \includegraphics[width=0.825\textwidth]{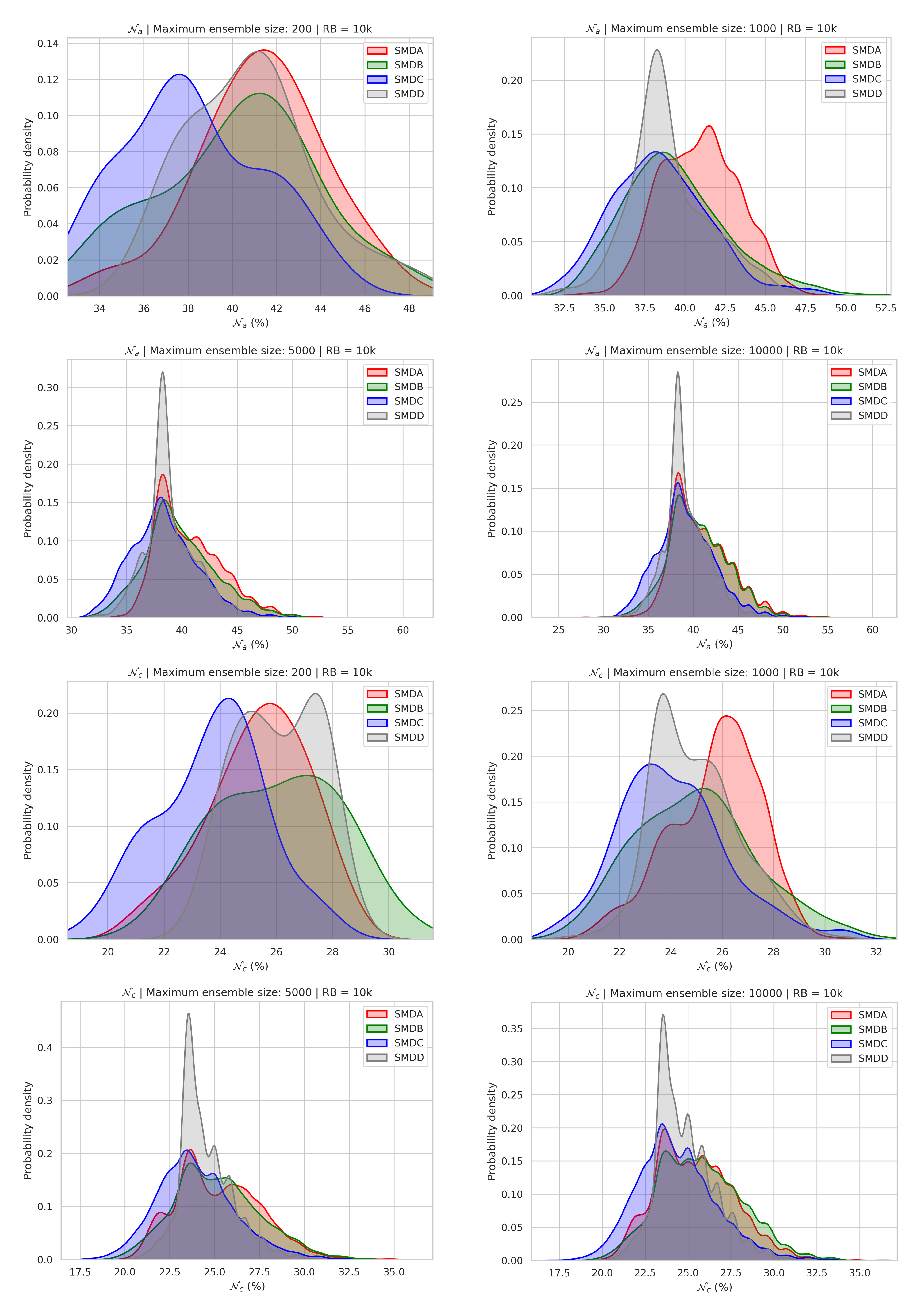}
    \caption{Complete \(\mathcal{N}_{a}\) and \(\mathcal{N}_{c}\) distributions for the Smad7 parameter sweep at RB = 10k (corresponding to the distributions summarized in Results 1, Figure 3b). Kernel density estimates are shown for \(\mathcal{N}_{a}\) (top four panels) and \(\mathcal{N}_{c}\) (bottom four panels) across the four Smad7 conformations (SMDA-SMDD; legend colors). Columns correspond to the maximum retained ensemble size \(n_{\mathrm{ret}}\in\{200,1000,5000,10000\}\).}
    \label{figS2}
\end{figure}

\newpage

%%%%%%%%%%%%%%%%%%

\begin{figure}[H]
    \centering
    \includegraphics[width=0.825\textwidth]{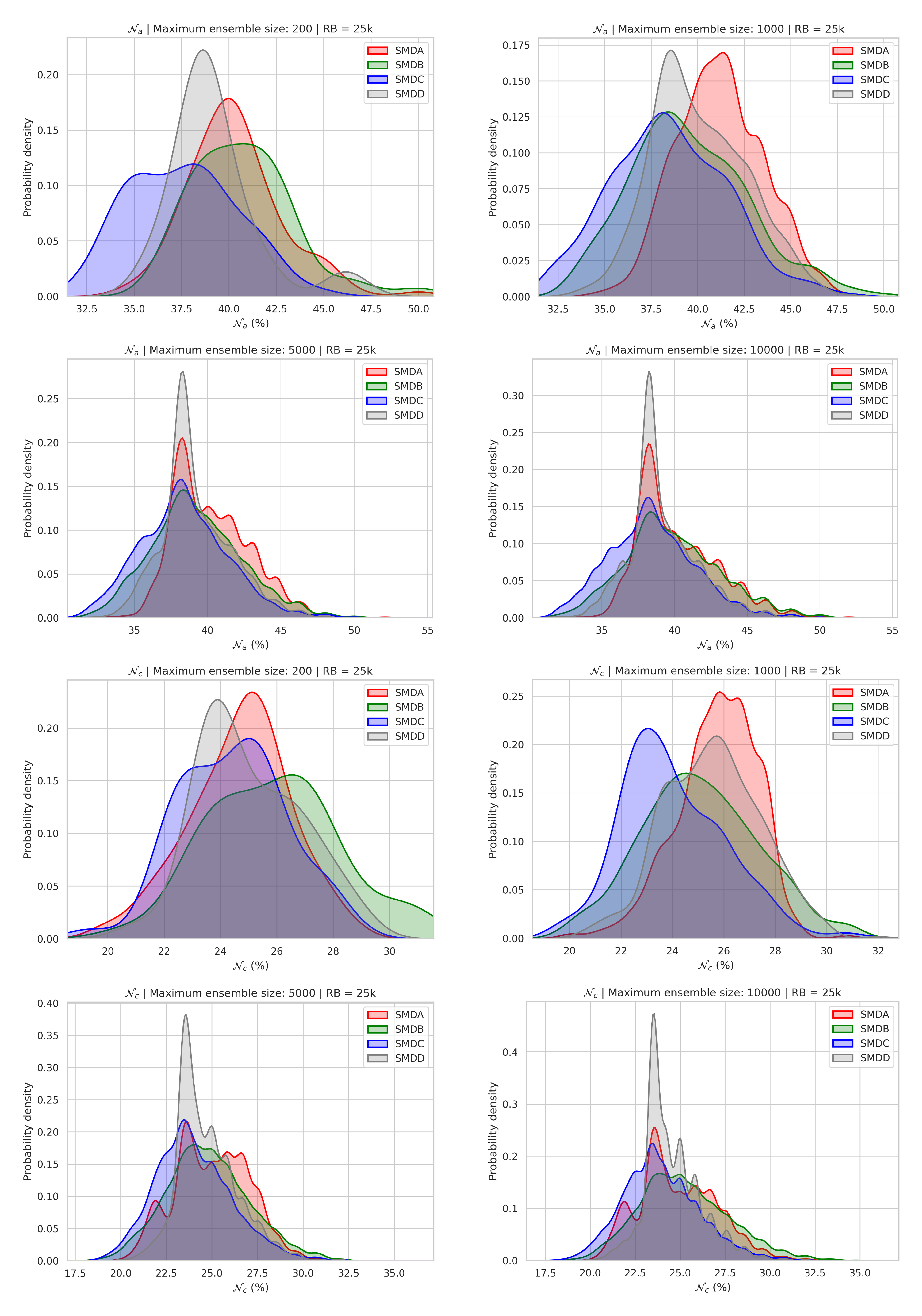}
    \caption{Complete \(\mathcal{N}_{a}\) and \(\mathcal{N}_{c}\) distributions for the Smad7 parameter sweep at RB = 25k (corresponding to the distributions summarized in Results 1, Figure 3b). Kernel density estimates are shown for \(\mathcal{N}_{a}\) (top four panels) and \(\mathcal{N}_{c}\) (bottom four panels) across the four Smad7 conformations (SMDA-SMDD; legend colors). Columns correspond to the maximum retained ensemble size \(n_{\mathrm{ret}}\in\{200,1000,5000,10000\}\).}
    \label{figS3}
\end{figure}

\newpage

%%%%%%%%%%%%%%%%%%

\begin{figure}[H]
\centering
    \includegraphics[width=0.82\textwidth]{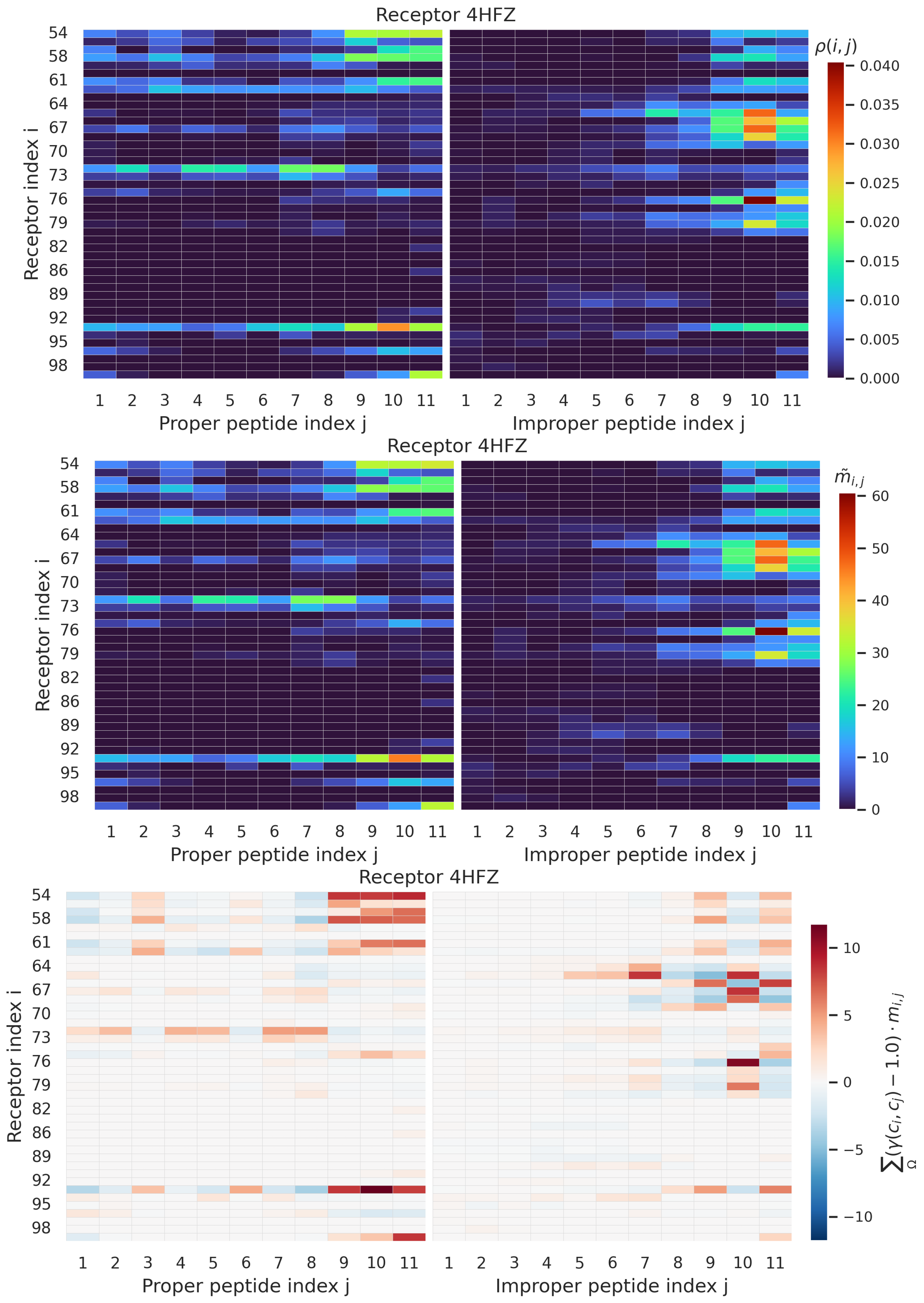}
    \caption{Contact map summaries for receptor 4HFZ under the cross-fertilization protocol. (top) Ensemble-averaged contact map probabilities, \(\rho(i,j)\). (middle) Contact-class weighted probability map, \(\tilde{m}_{i,j}\), reflecting the enthalpic contribution of individual contacts. (bottom) The ensemble-level net contact-class modulation, \(\sum_{\Omega} (\gamma(c_i, c_j) - 1.0) \cdot m_{i,j}\), where positive (red) and negative (blue) values denote regions of favorable and unfavorable biophysical modulation respectively.}
    \label{figS4}
\end{figure}

\newpage

%%%%%%%%%%%%%%%%%%

\begin{figure}[H]
    \centering
    \includegraphics[width=0.82\textwidth]{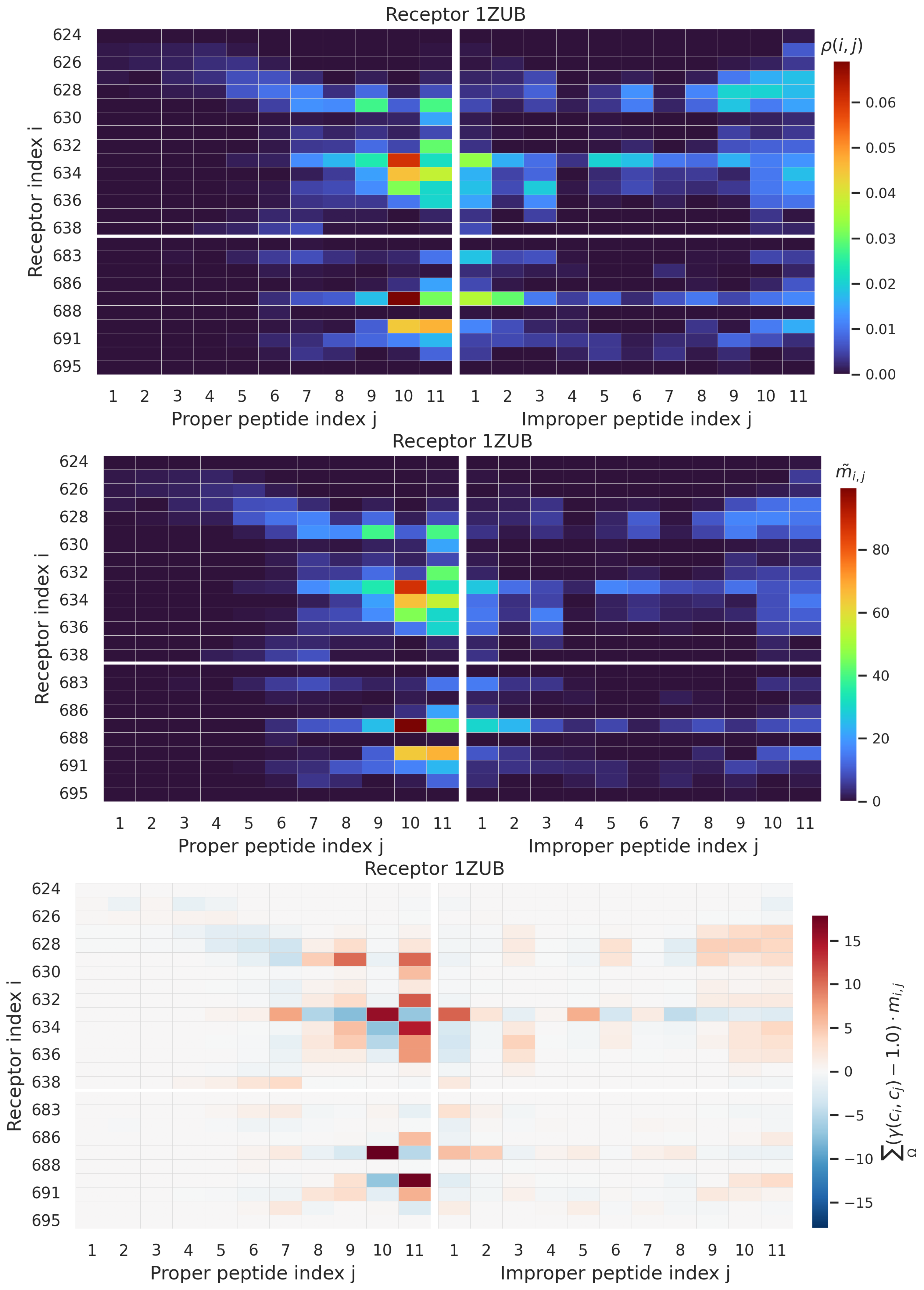}
    \caption{Contact map summaries for receptor 1ZUB under the cross-fertilization protocol. (top) Ensemble-averaged contact map probabilities, \(\rho(i,j)\). (middle) Contact-class weighted probability map, \(\tilde{m}_{i,j}\), reflecting the enthalpic contribution of individual contacts. (bottom) The ensemble-level net contact-class modulation, \(\sum_{\Omega} (\gamma(c_i, c_j) - 1.0) \cdot m_{i,j}\), where positive (red) and negative (blue) values denote regions of favorable and unfavorable biophysical modulation respectively. For 1ZUB, the active residues are discontinuous in sequence shown by a white horizontal space between residues 638 and 683.}
    \label{figS5}
\end{figure}

\newpage

%%%%%%%%%%%%%%%%%%

\begin{figure}[H]
    \centering
    \includegraphics[width=1.0\textwidth]{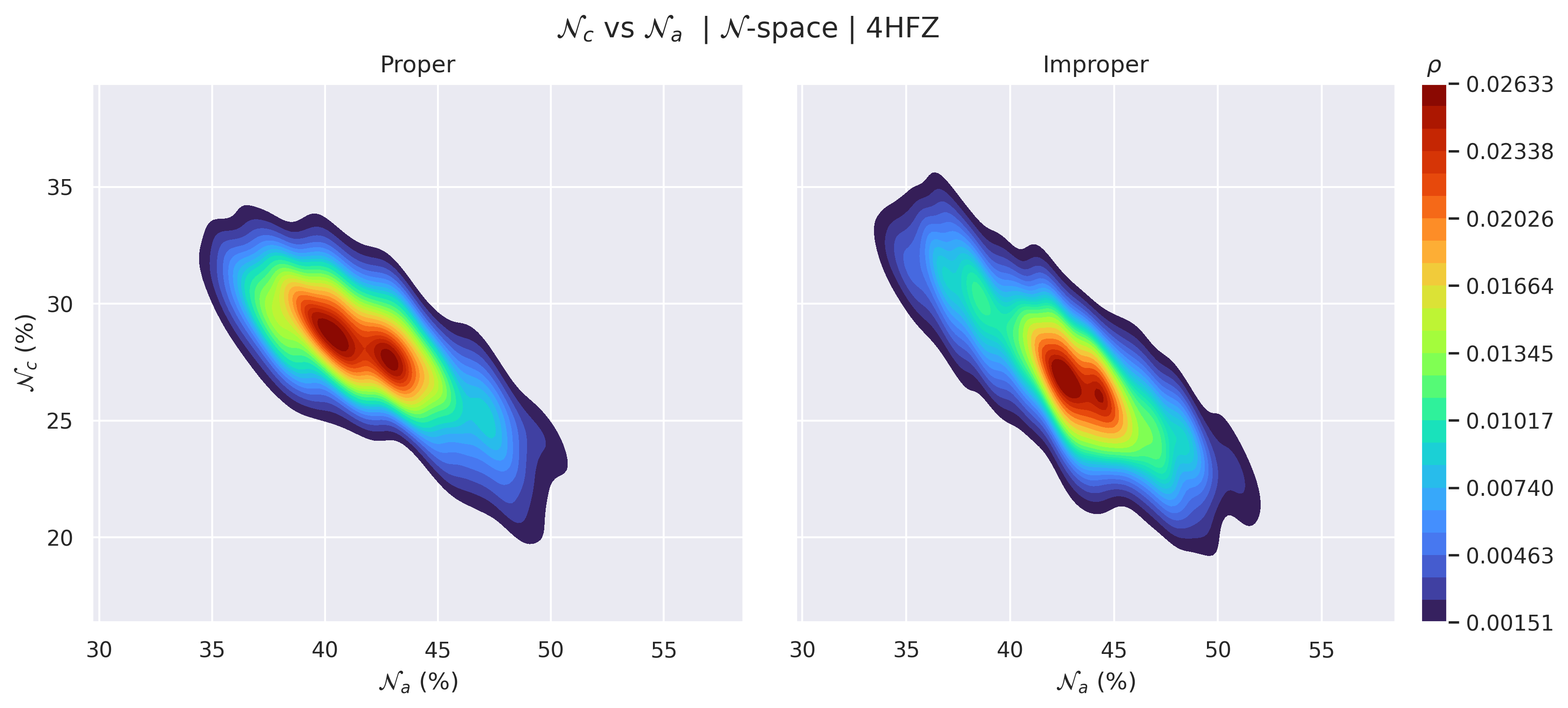}
    \caption{Phase-pooled bivariate \(\mathcal{N}\)-space density for receptor 4HFZ under the cross-fertilization protocol (proper vs improper). Microstates are pooled across replicas within each phase.}
    \label{figS6}
\end{figure}

%%%%%%%%%%%%%%%%%%

\begin{figure}[H]
    \centering
    \includegraphics[width=1.0\textwidth]{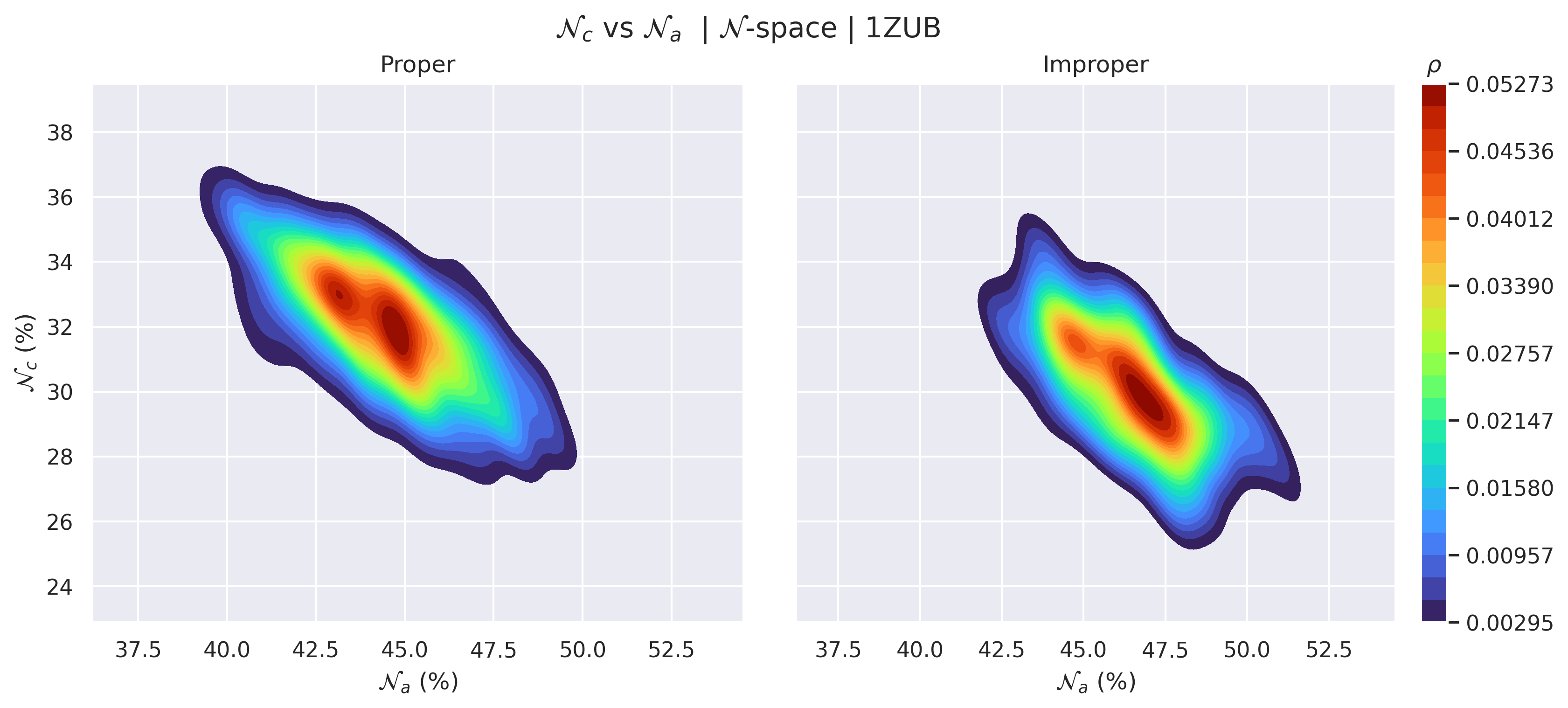}
    \caption{Phase-pooled bivariate \(\mathcal{N}\)-space density for receptor 1ZUB under the cross-fertilization protocol (proper vs improper). Microstates are pooled across replicas within each phase.}
    \label{figS7}
\end{figure}

\newpage

%%%%%%%%%%%%%%%%%%%%%%%%

\begin{figure}[H]
    \centering
    \includegraphics[width=0.62\textwidth]{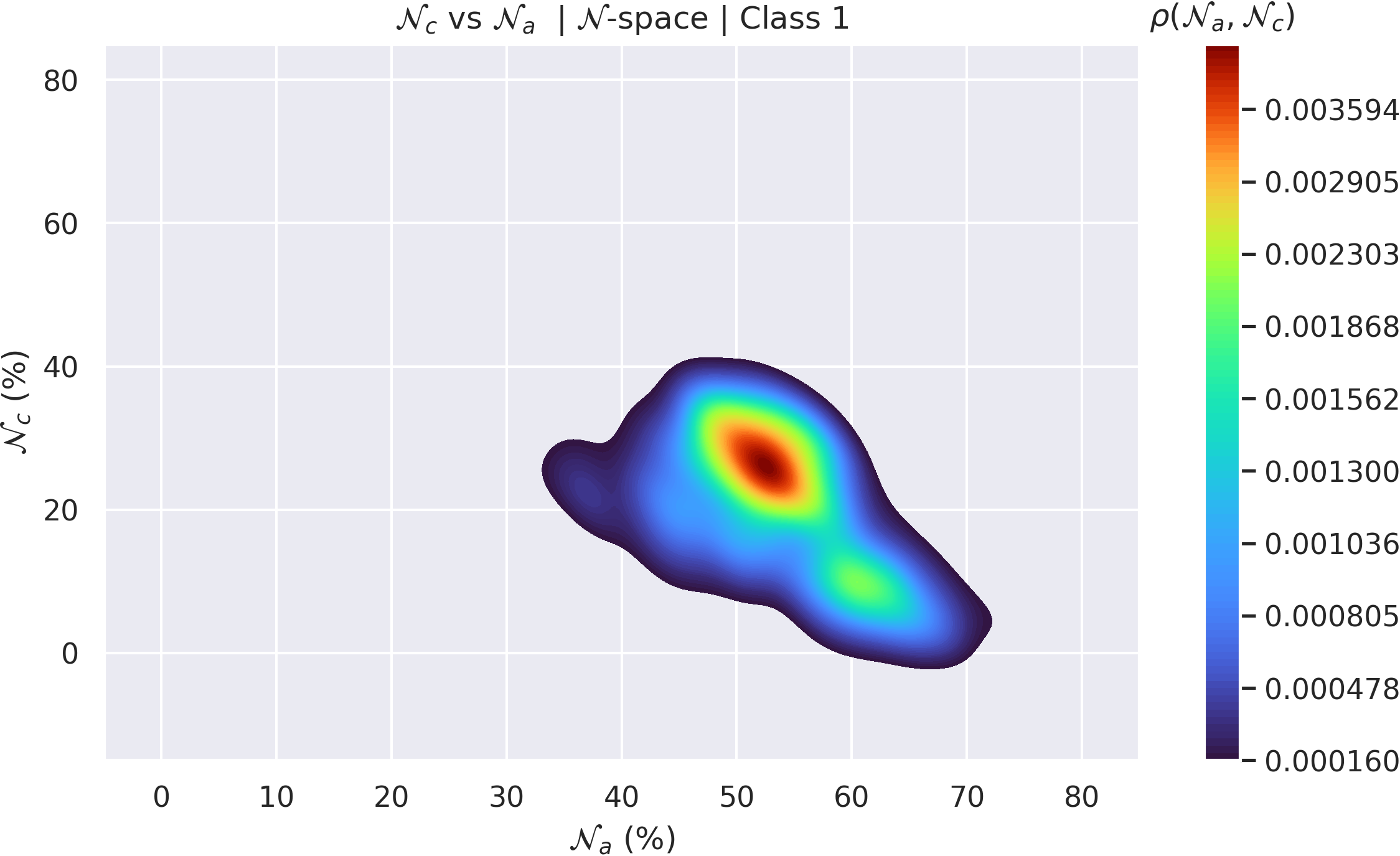} \\
    \includegraphics[width=0.62\textwidth]{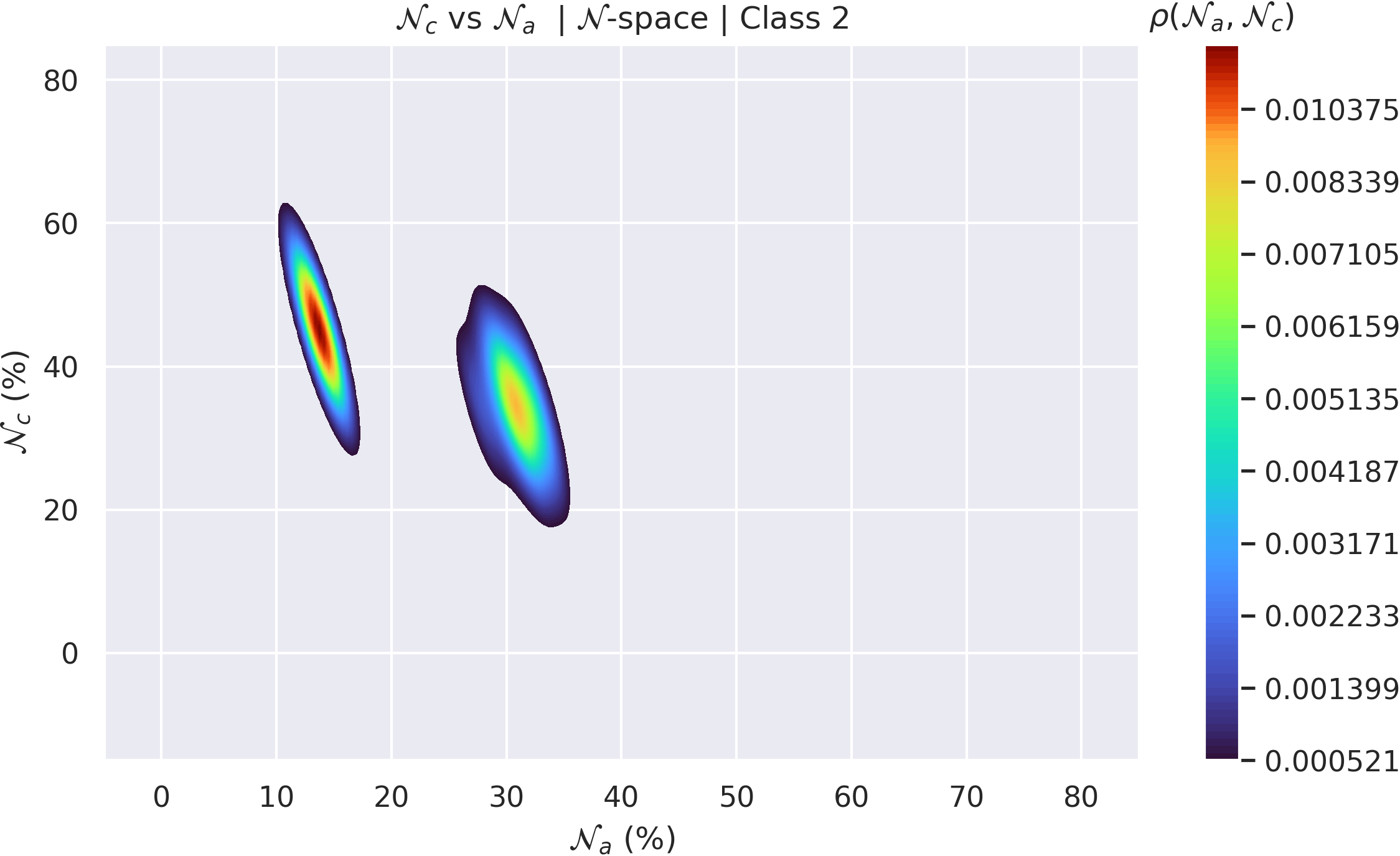} \\
    \includegraphics[width=0.62\textwidth]{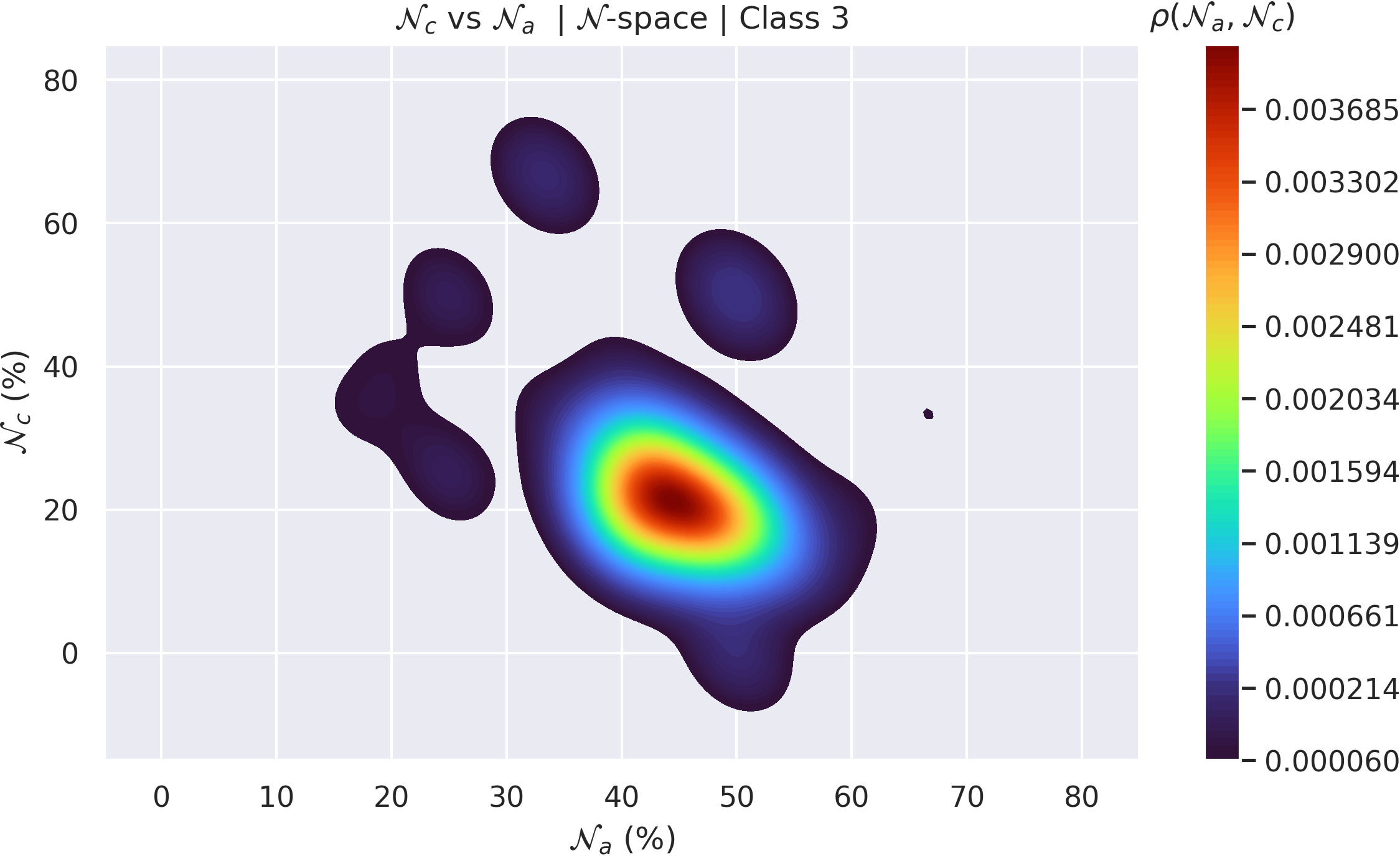}
    \caption{Phase-pooled bivariate \(\mathcal{N}\)-space density for the three experimental meta-ensemble classes: (top) Class 1, (middle) Class 2, and (bottom) Class 3. Densities are computed using the RSA-thresholded count metric with explicit interface exclusion.}
    \label{figS8}
\end{figure}

\phantomsection
\label{supp:figures:end}

\newpage

%%%%%%%%%%%%%%%%%%%%%%%%

%%%%%%%%%%%%%%%%%%%%%%%%%%%%%%%%%%%%%%%%%%%%
% ---------- Supporting Tables ----------

\section*{Supporting Tables}
\phantomsection
\label{supp:tables:start}

\begin{table}[h]
\centering
\caption{NACCESS residue maximum solvent-accessible surface area values (\(\mathrm{ASA}_{\max}\)) for SASA to RSA calculations.}
\label{tableS1}
\setlength{\tabcolsep}{2pt}
{\renewcommand{\arraystretch}{1.08}
\begin{tabular}{|c|c|}
\hline
Residue & \(\mathrm{ASA}_{\max}\) \\
\hline
Alanine (ALA) & 107.95\,\AA$^{2}$ \\ \hline
Cysteine (CYS) & 134.28\,\AA$^{2}$ \\ \hline
Aspartic acid (ASP) & 140.39\,\AA$^{2}$ \\ \hline
Glutamic acid (GLU) & 172.25\,\AA$^{2}$ \\ \hline
Phenylalanine (PHE) & 199.48\,\AA$^{2}$ \\ \hline
Glycine (GLY) & 80.10\,\AA$^{2}$ \\ \hline
Histidine (HIS) & 182.88\,\AA$^{2}$ \\ \hline
Isoleucine (ILE) & 175.12\,\AA$^{2}$ \\ \hline
Lysine (LYS) & 200.81\,\AA$^{2}$ \\ \hline
Leucine (LEU) & 178.63\,\AA$^{2}$ \\ \hline
Methionine (MET) & 194.15\,\AA$^{2}$ \\ \hline
Asparagine (ASN) & 143.94\,\AA$^{2}$ \\ \hline
Proline (PRO) & 136.13\,\AA$^{2}$ \\ \hline
Glutamine (GLN) & 178.50\,\AA$^{2}$ \\ \hline
Arginine (ARG) & 238.76\,\AA$^{2}$ \\ \hline
Serine (SER) & 116.50\,\AA$^{2}$ \\ \hline
Threonine (THR) & 139.27\,\AA$^{2}$ \\ \hline
Valine (VAL) & 151.44\,\AA$^{2}$ \\ \hline
Tryptophan (TRP) & 249.36\,\AA$^{2}$ \\ \hline
Tyrosine (TYR) & 212.76\,\AA$^{2}$ \\
\hline
\end{tabular}
}
\end{table}

\begin{table}[h]
\centering
\caption{Pairwise chemical weights for residue-residue contact classes.}
\label{tableS2}
\setlength{\tabcolsep}{2pt}
\resizebox{0.8\textwidth}{!}
{
\begin{tabular}{|c|c|c|}
\hline
Contact Class & $\gamma(c_i,c_j)$ & Note \\ \hline
positive-negative & 1.65 & Salt-bridge attraction, strongly favorable \\ \hline
negative-positive & 1.65 & Salt-bridge attraction, strongly favorable \\ \hline
apolar-apolar     & 1.35 & Hydrophobic contacts, drives concentration, moderately favorable \\ \hline
polar-polar       & 1.22 & H-bond/dipole interactions, moderately favorable \\ \hline
polar-positive    & 1.22 & Polar assists charge via H-bond/electrostatics, moderately favorable \\ \hline
polar-negative    & 1.22 & Polar assists charge via H-bond/electrostatics, moderately favorable \\ \hline
apolar-polar      & 0.90 & Induces desolvation costs, mildly unfavorable \\ \hline
apolar-positive   & 0.82 & Hydrophobe-charge mismatch, moderately unfavorable \\ \hline
apolar-negative   & 0.82 & Hydrophobe-charge mismatch, moderately unfavorable \\ \hline
positive-positive & 0.61 & Like-charge repulsion, strongly unfavorable \\ \hline
negative-negative & 0.61 & Like-charge repulsion, strongly unfavorable \\ \hline
\end{tabular}
}
\end{table}

\newpage

%%%%%%%%%%%%%%%%%%%%%%%%%

\begin{table}[h]
\centering
\caption{List of \{PY\} (PPxY, LPxY, and PPGW) protein sequences.}
\label{tableS3}
\setlength{\tabcolsep}{2pt}
\begin{tabular}{|c|c|c|}
\hline
PDB ID & Sequence & Length \\ \hline
1ZCN & LPPGWEKRMS & 10 \\ \hline
2LB1 & PPAYLPPEDP & 10 \\ \hline
2LB2 & PPPGYLSEDG & 10 \\ \hline
2LTV & PPPYSRYPMD & 10 \\ \hline
2ZQS & KLPPGWEKRM & 10 \\ \hline
2EZ5 & TGLPSYDEALH & 11 \\ \hline
2KXQ & SPPPPYSRYPM & 11 \\ \hline
6J1Z & PPGWEKRTDPR & 11 \\ \hline
7BQF & LPLPPGWSVDW & 11 \\ \hline
2LAW & TPPPAYLPPEDP & 12 \\ \hline
2LTY & ESPPPPYSRYPMD & 13 \\ \hline
2MPT & RLDLPPYETFEDL & 13 \\ \hline
7LP5a & QNNEELPTYEEAKV & 14 \\ \hline
SMDA & GESPPPPYSRYPMD & 14 \\ \hline
SMDC & GESPPPPYSRYPMD & 14 \\ \hline
2DJY & ELESPPPPYSRYPMD & 15 \\ \hline
\end{tabular}
\end{table}

\begin{table}[h]
\centering
\caption{List of randomly generated protein sequences in the \{RD\} set.}
\label{tableS4}
\setlength{\tabcolsep}{2pt}
\begin{tabular}{|c|c|c|}
\hline
Name & Sequence & Length \\ \hline
RD01 & YSEEITHQQM & 10 \\ \hline
RD02 & RHAKHWMSRG & 10 \\ \hline
RD03 & FWISKKCTKW & 10 \\ \hline
RD04 & HATCCYDVKDF & 11 \\ \hline
RD05 & FWEVFMGRLLE & 11 \\ \hline
RD06 & MQQMGEFKKNW & 11 \\ \hline
RD07 & EDAEIYEPYFNW & 12 \\ \hline
RD08 & CFPYYGYTTIMH & 12 \\ \hline
RD09 & MMIRYLHNVKSDW & 13 \\ \hline
RD10 & DFNTHYGVVSIET & 13 \\ \hline
RD11 & SIVFMYTLMFMMIN & 14 \\ \hline
RD12 & PIHLDARAPRYWQN & 14 \\ \hline
RD13 & KNQFLMIWHAYYIL & 14 \\ \hline
RD14 & VMWECAYSNTHWSGI & 15 \\ \hline
RD15 & YDVLQQVSFMTNYWI & 15 \\ \hline
RD16 & SYYSFWDGRIHQTGA & 15 \\ \hline
\end{tabular}
\end{table}

\newpage

%%%%%%%%%%%%%%%%%%%%%%%%

\begingroup
\setlength{\intextsep}{6pt}
\setlength{\textfloatsep}{6pt}
\setlength{\floatsep}{6pt}

\begin{table}[H]
\caption{$\mathcal{N}$-state properties between Step 2 (semi-flexible refinement) and Step 3 (MDS in explicit solvent) extracted from conformational ensembles (docking solutions). Sixteen proline-rich peptides with motifs PPxY, LPxY, and PPGW were used to generate ensembles of protein-peptide complexes. The second column from the left contains the variables $\mathcal{N}_a$ and $\mathcal{N}_c$ being the NIS apolar and charged surface percentages respectively. $P_{max}$ represents the maximum probability state (mode) in $\mathcal{N}$-space, and $S_{\Psi}$ is the peripheral surface information (PSI) entropy for a given ensemble. The terms $\Delta \mathcal{N}$ and $\Delta S_{\Psi}$ indicate the changes from Step 2 to Step 3 in the number of $\mathcal{N}$-states and PSI entropy respectively.}
\label{tableS5}
\setlength{\tabcolsep}{2pt}
\resizebox{1.0\textwidth}{!}
{
\renewcommand{\arraystretch}{1.12}
\begin{tabular}{|cc|ccccccccc|ccc|cccc|}
\hline
 &  &  &  &  &  & PPxY &  &  &  &  &  & LPxY &  &  & \multicolumn{2}{c}{PPGW} &  \\ \cline{3-18} 
 &  & \multicolumn{1}{c|}{\small 2DJY} & \multicolumn{1}{c|}{\small 2KXQ} & \multicolumn{1}{c|}{\small 2LB1} & \multicolumn{1}{c|}{\small 2LB2} & \multicolumn{1}{c|}{\small 2LTV} & \multicolumn{1}{c|}{\small 2LTY} & \multicolumn{1}{c|}{\small 2LAW} & \multicolumn{1}{c|}{\small SMDA} & \small SMDC & \multicolumn{1}{c|}{\small 2EZ5} & \multicolumn{1}{c|}{\small 7LP5a} & \small 2MPT & \multicolumn{1}{c|}{\small 1ZCN} & \multicolumn{1}{c|}{\small 2ZQS} & \multicolumn{1}{c|}{\small 6JIZ} & \small 7BQF \\ \hline
\multicolumn{1}{|c|}{\multirow{5}{*}{{\rotatebox{90}{\hspace{-0.3cm} Step 2}}}} & \small $\mathcal{N}_{a}$ (\%) & \multicolumn{1}{c|}{40.00} & \multicolumn{1}{c|}{41.30} & \multicolumn{1}{c|}{44.44} & \multicolumn{1}{c|}{42.22} & \multicolumn{1}{c|}{40.00} & \multicolumn{1}{c|}{39.58} & \multicolumn{1}{c|}{44.68} & \multicolumn{1}{c|}{40.82} & 40.00 & \multicolumn{1}{c|}{39.13} & \multicolumn{1}{c|}{34.69} & 39.58 & \multicolumn{1}{c|}{40.00} & \multicolumn{1}{c|}{40.00} & \multicolumn{1}{c|}{36.96} & 43.48 \\ \cline{2-18} 
\multicolumn{1}{|c|}{} & \small $\mathcal{N}_{c}$ (\%) & \multicolumn{1}{c|}{24.00} & \multicolumn{1}{c|}{19.57} & \multicolumn{1}{c|}{22.22} & \multicolumn{1}{c|}{22.22} & \multicolumn{1}{c|}{22.22} & \multicolumn{1}{c|}{22.92} & \multicolumn{1}{c|}{21.28} & \multicolumn{1}{c|}{22.45} & 24.00 & \multicolumn{1}{c|}{21.74} & \multicolumn{1}{c|}{26.53} & 27.08 & \multicolumn{1}{c|}{24.44} & \multicolumn{1}{c|}{26.67} & \multicolumn{1}{c|}{28.26} & 19.57 \\ \cline{2-18} 
\multicolumn{1}{|c|}{} & \small $P_{max}$ & \multicolumn{1}{c|}{0.80} & \multicolumn{1}{c|}{0.56} & \multicolumn{1}{c|}{0.88} & \multicolumn{1}{c|}{0.86} & \multicolumn{1}{c|}{0.91} & \multicolumn{1}{c|}{0.70} & \multicolumn{1}{c|}{0.80} & \multicolumn{1}{c|}{0.79} & 0.80 & \multicolumn{1}{c|}{0.62} & \multicolumn{1}{c|}{0.76} & 0.87 & \multicolumn{1}{c|}{0.43} & \multicolumn{1}{c|}{0.93} & \multicolumn{1}{c|}{0.43} & 0.88 \\ \cline{2-18} 
\multicolumn{1}{|c|}{} & \small $\mathcal{N}$-states & \multicolumn{1}{c|}{4.00} & \multicolumn{1}{c|}{7.00} & \multicolumn{1}{c|}{4.00} & \multicolumn{1}{c|}{6.00} & \multicolumn{1}{c|}{3.00} & \multicolumn{1}{c|}{5.00} & \multicolumn{1}{c|}{6.00} & \multicolumn{1}{c|}{4.00} & 4.00 & \multicolumn{1}{c|}{9.00} & \multicolumn{1}{c|}{6.00} & 3.00 & \multicolumn{1}{c|}{14.00} & \multicolumn{1}{c|}{4.00} & \multicolumn{1}{c|}{6.00} & 3.00 \\ \cline{2-18} 
\multicolumn{1}{|c|}{} & \small $S_{\Psi}$ & \multicolumn{1}{c|}{-61.23} & \multicolumn{1}{c|}{-43.94} & \multicolumn{1}{c|}{-76.27} & \multicolumn{1}{c|}{-50.27} & \multicolumn{1}{c|}{-242.69} & \multicolumn{1}{c|}{-29.28} & \multicolumn{1}{c|}{-69.30} & \multicolumn{1}{c|}{-85.27} & -61.23 & \multicolumn{1}{c|}{-22.70} & \multicolumn{1}{c|}{-37.29} & -123.54 & \multicolumn{1}{c|}{-7.32} & \multicolumn{1}{c|}{-73.41} & \multicolumn{1}{c|}{-44.42} & -76.43 \\ \hline
\multicolumn{1}{|c|}{\multirow{5}{*}{{\rotatebox{90}{\hspace{-0.3cm} Step 3}}}} & \small $\mathcal{N}_{a}$ (\%) & \multicolumn{1}{c|}{40.00} & \multicolumn{1}{c|}{41.30} & \multicolumn{1}{c|}{44.44} & \multicolumn{1}{c|}{42.22} & \multicolumn{1}{c|}{40.00} & \multicolumn{1}{c|}{39.58} & \multicolumn{1}{c|}{44.68} & \multicolumn{1}{c|}{40.82} & 40.00 & \multicolumn{1}{c|}{39.13} & \multicolumn{1}{c|}{34.69} & 39.58 & \multicolumn{1}{c|}{40.91} & \multicolumn{1}{c|}{40.00} & \multicolumn{1}{c|}{36.96} & 43.48 \\ \cline{2-18} 
\multicolumn{1}{|c|}{} & \small $\mathcal{N}_{a}$ (\%) & \multicolumn{1}{c|}{24.00} & \multicolumn{1}{c|}{19.57} & \multicolumn{1}{c|}{22.22} & \multicolumn{1}{c|}{22.22} & \multicolumn{1}{c|}{22.22} & \multicolumn{1}{c|}{22.92} & \multicolumn{1}{c|}{21.28} & \multicolumn{1}{c|}{22.45} & 24.00 & \multicolumn{1}{c|}{21.74} & \multicolumn{1}{c|}{26.53} & 27.08 & \multicolumn{1}{c|}{25.00} & \multicolumn{1}{c|}{26.67} & \multicolumn{1}{c|}{28.26} & 19.57 \\ \cline{2-18} 
\multicolumn{1}{|c|}{} & \small $P_{max}$ & \multicolumn{1}{c|}{0.68} & \multicolumn{1}{c|}{0.50} & \multicolumn{1}{c|}{0.69} & \multicolumn{1}{c|}{0.74} & \multicolumn{1}{c|}{0.79} & \multicolumn{1}{c|}{0.54} & \multicolumn{1}{c|}{0.60} & \multicolumn{1}{c|}{0.69} & 0.68 & \multicolumn{1}{c|}{0.46} & \multicolumn{1}{c|}{0.53} & 0.64 & \multicolumn{1}{c|}{0.32} & \multicolumn{1}{c|}{0.69} & \multicolumn{1}{c|}{0.29} & 0.88 \\ \cline{2-18} 
\multicolumn{1}{|c|}{} & \small $\mathcal{N}$-states & \multicolumn{1}{c|}{8.00} & \multicolumn{1}{c|}{9.00} & \multicolumn{1}{c|}{7.00} & \multicolumn{1}{c|}{8.00} & \multicolumn{1}{c|}{6.00} & \multicolumn{1}{c|}{10.00} & \multicolumn{1}{c|}{7.00} & \multicolumn{1}{c|}{7.00} & 8.00 & \multicolumn{1}{c|}{12.00} & \multicolumn{1}{c|}{8.00} & 4.00 & \multicolumn{1}{c|}{12.00} & \multicolumn{1}{c|}{7.00} & \multicolumn{1}{c|}{10.00} & 4.00 \\ \cline{2-18} 
\multicolumn{1}{|c|}{} & \small $S_{\Psi}$ & \multicolumn{1}{c|}{-20.10} & \multicolumn{1}{c|}{-24.65} & \multicolumn{1}{c|}{-27.74} & \multicolumn{1}{c|}{-27.64} & \multicolumn{1}{c|}{-83.38} & \multicolumn{1}{c|}{-7.62} & \multicolumn{1}{c|}{-44.88} & \multicolumn{1}{c|}{-41.78} & -20.10 & \multicolumn{1}{c|}{-9.10} & \multicolumn{1}{c|}{-13.23} & -68.95 & \multicolumn{1}{c|}{-9.12} & \multicolumn{1}{c|}{-22.57} & \multicolumn{1}{c|}{-13.72} & -30.79 \\ \hline
\multicolumn{1}{|c|}{} & \small $\Delta \mathcal{N}$ & \multicolumn{1}{c|}{4.00} & \multicolumn{1}{c|}{2.00} & \multicolumn{1}{c|}{3.00} & \multicolumn{1}{c|}{2.00} & \multicolumn{1}{c|}{3.00} & \multicolumn{1}{c|}{5.00} & \multicolumn{1}{c|}{1.00} & \multicolumn{1}{c|}{3.00} & 4.00 & \multicolumn{1}{c|}{3.00} & \multicolumn{1}{c|}{2.00} & 1.00 & \multicolumn{1}{c|}{-2.00} & \multicolumn{1}{c|}{3.00} & \multicolumn{1}{c|}{4.00} & 1.00 \\ \cline{2-18} 
\multicolumn{1}{|c|}{} & \small $\Delta S_{\Psi}$ & \multicolumn{1}{c|}{41.13} & \multicolumn{1}{c|}{19.29} & \multicolumn{1}{c|}{48.53} & \multicolumn{1}{c|}{22.63} & \multicolumn{1}{c|}{159.31} & \multicolumn{1}{c|}{21.66} & \multicolumn{1}{c|}{24.42} & \multicolumn{1}{c|}{43.49} & 41.13 & \multicolumn{1}{c|}{13.60} & \multicolumn{1}{c|}{24.05} & 54.59 & \multicolumn{1}{c|}{-1.79} & \multicolumn{1}{c|}{50.84} & \multicolumn{1}{c|}{30.70} & 45.63 \\ \hline
\end{tabular}
}
\end{table}

%\newpage

%%%%%%%%%%%%%%%%%%%%%%%%

\vspace{0.5cm}

\begin{table}[H]
\caption{$\mathcal{N}$-state properties between Step 2 (semi-flexible refinement) and Step 3 (MDS in explicit solvent) extracted from conformational ensembles (docking solutions). Sixteen random sequences (decoys) were predicted using AF3 then equilibrated with MDS in explicit TIP3P solvent (RD01 to RD16). The second column contains the variables $\mathcal{N}_a$ and $\mathcal{N}_c$ which are the NIS apolar and charged surface percentages respectively. $P_{max}$ represents the maximum probability state (mode) in $\mathcal{N}$-space, and $S_{\Psi}$ is the peripheral surface information (PSI) entropy for a given conformational ensemble. The terms $\Delta \mathcal{N}$ and $\Delta S_{\Psi}$ indicate the changes from Step 2 to Step 3 in the number of $\mathcal{N}$-states and PSI entropy respectively.}
\label{tableS6}
\setlength{\tabcolsep}{2pt}
\resizebox{1.0\textwidth}{!}
{
\renewcommand{\arraystretch}{1.12}
\begin{tabular}{|cc|c|c|c|c|c|c|c|c|c|c|c|c|c|c|c|c|}
\hline
 &  & \small RD01 & \small RD02 & \small RD03 & \small RD04 & \small RD05 & \small RD06 & \small RD07 & \small RD08 & \small RD09 & \small RD10 & \small RD11 & \small RD12 & \small RD13 & \small RD14 & \small RD15 & \small RD16 \\ \hline
\multicolumn{1}{|c|}{\multirow{5}{*}{{\rotatebox{90}{\hspace{-0.3cm} Step 2}}}} & \small $\mathcal{N}_{a}$ (\%) & 33.33 & 35.56 & 33.33 & 34.78 & 43.48 & 36.96 & 36.17 & 38.30 & 37.50 & 37.50 & 46.94 & 38.78 & 40.82 & 36.00 & 38.00 & 36.00 \\ \cline{2-18} 
\multicolumn{1}{|c|}{} & \small $\mathcal{N}_{c}$ (\%) & 22.22 & 24.44 & 24.44 & 23.91 & 23.91 & 23.91 & 25.53 & 17.02 & 22.92 & 20.83 & 16.33 & 22.45 & 18.37 & 18.00 & 18.00 & 20.00 \\ \cline{2-18} 
\multicolumn{1}{|c|}{} & \small $P_{max}$ & 0.92 & 0.63 & 0.89 & 0.69 & 0.83 & 0.95 & 0.71 & 0.58 & 0.67 & 0.67 & 0.82 & 0.76 & 0.70 & 0.94 & 0.87 & 0.90 \\ \cline{2-18} 
\multicolumn{1}{|c|}{} & \small $\mathcal{N}$-states & 4.00 & 7.00 & 4.00 & 7.00 & 5.00 & 2.00 & 5.00 & 6.00 & 6.00 & 6.00 & 3.00 & 5.00 & 7.00 & 2.00 & 5.00 & 5.00 \\ \cline{2-18} 
\multicolumn{1}{|c|}{} & \small $S_{\Psi}$ & -142.64 & -33.60 & -117.95 & -23.97 & -70.79 & -151.29 & -79.61 & -34.26 & -19.42 & -44.17 & -24.71 & -35.29 & -9.68 & -140.52 & -27.13 & -81.42 \\ \hline
\multicolumn{1}{|c|}{\multirow{5}{*}{{\rotatebox{90}{\hspace{-0.3cm} Step 3}}}} & \small $\mathcal{N}_{a}$ (\%) & 33.33 & 35.56 & 33.33 & 34.78 & 43.48 & 36.96 & 36.17 & 38.30 & 37.50 & 37.50 & 46.94 & 38.78 & 40.82 & 36.00 & 38.00 & 36.00 \\ \cline{2-18} 
\multicolumn{1}{|c|}{} & \small $\mathcal{N}_{a}$ (\%) & 22.22 & 24.44 & 24.44 & 23.91 & 23.91 & 23.91 & 25.53 & 17.02 & 22.92 & 20.83 & 16.33 & 22.45 & 18.37 & 18.00 & 18.00 & 20.00 \\ \cline{2-18} 
\multicolumn{1}{|c|}{} & \small $P_{max}$ & 0.73 & 0.52 & 0.71 & 0.46 & 0.76 & 0.83 & 0.57 & 0.38 & 0.56 & 0.48 & 0.49 & 0.66 & 0.67 & 0.70 & 0.70 & 0.72 \\ \cline{2-18} 
\multicolumn{1}{|c|}{} & \small $\mathcal{N}$-states & 6.00 & 9.00 & 6.00 & 11.00 & 4.00 & 6.00 & 8.00 & 8.00 & 9.00 & 8.00 & 8.00 & 5.00 & 10.00 & 6.00 & 7.00 & 7.00 \\ \cline{2-18} 
\multicolumn{1}{|c|}{} & \small $S_{\Psi}$ & -67.30 & -18.22 & -60.30 & -7.79 & -87.38 & -32.19 & -30.90 & -15.71 & -5.19 & -18.71 & -3.57 & -34.00 & -5.21 & -26.59 & -11.50 & -40.12 \\ \hline
\multicolumn{1}{|c|}{} & \small $\Delta \mathcal{N}$ & 2.00 & 2.00 & 2.00 & 4.00 & -1.00 & 4.00 & 3.00 & 2.00 & 3.00 & 2.00 & 2.00 & 0.00 & 3.00 & 4.00 & 2.00 & 2.00 \\ \cline{2-18} 
\multicolumn{1}{|c|}{} & \small $\Delta S_{\Psi}$ & 75.34 & 15.38 & 57.65 & 16.17 & -16.59 & 119.10 & 48.71 & 18.56 & 14.23 & 25.46 & 21.14 & 1.30 & 4.47 & 113.93 & 15.63 & 41.31 \\ \hline
\end{tabular}
}
\end{table}

%%%%%%%%%%%%%%%%%%%%%%%%
%%%%%%%%%%%%%%%%%%%%%%%%

\endgroup

\begin{table}[H]
\centering
\caption{Experimental meta-ensemble systems used in the third Results (experimental WW domains). Year denotes the initial PDB release year, the number of conformers in the NMR ensembles are listed, and resolution is reported only for X-ray diffraction entries.}
\label{tableS7}
\setlength{\tabcolsep}{4pt}
{\small
\renewcommand{\arraystretch}{1.20}
\begin{tabular}{|c|c|c|c|c|}
\hline
PDB ID & \# conformers & Experimental method & Resolution (\AA) & Year \\ \hline
1I5H & 15 & Solution NMR & - & 2001 \\ \hline
1I8G & 10 & Solution NMR & - & 2001 \\ \hline
1I8H & 10 & Solution NMR & - & 2001 \\ \hline
1JMQ & 20 & Solution NMR & - & 2001 \\ \hline
1K5R & 10 & Solution NMR & - & 2001 \\ \hline
1K9Q & 20 & Solution NMR & - & 2001 \\ \hline
1K9R & 20 & Solution NMR & - & 2001 \\ \hline
2DJY & 30 & Solution NMR & - & 2006 \\ \hline
2DYF & 20 & Solution NMR & - & 2006 \\ \hline
2EZ5 & 30 & Solution NMR & - & 2006 \\ \hline
2JMF & 20 & Solution NMR & - & 2007 \\ \hline
2JO9 & 10 & Solution NMR & - & 2007 \\ \hline
2JUP & 10 & Solution NMR & - & 2007 \\ \hline
2KQ0 & 20 & Solution NMR & - & 2010 \\ \hline
2LAJ & 15 & Solution NMR & - & 2011 \\ \hline
2LAW & 20 & Solution NMR & - & 2011 \\ \hline
2LAX & 20 & Solution NMR & - & 2011 \\ \hline
2LAY & 20 & Solution NMR & - & 2011 \\ \hline
2LAZ & 24 & Solution NMR & - & 2011 \\ \hline
2LB0 & 21 & Solution NMR & - & 2011 \\ \hline
2LB1 & 20 & Solution NMR & - & 2011 \\ \hline
2LB2 & 25 & Solution NMR & - & 2011 \\ \hline
2LB3 & 20 & Solution NMR & - & 2011 \\ \hline
2LTV & 40 & Solution NMR & - & 2012 \\ \hline
2LTW & 30 & Solution NMR & - & 2012 \\ \hline
2LTX & 25 & Solution NMR & - & 2012 \\ \hline
2LTY & 30 & Solution NMR & - & 2012 \\ \hline
2LTZ & 20 & Solution NMR & - & 2012 \\ \hline
2M3O & 15 & Solution NMR & - & 2013 \\ \hline
2MPT & 20 & Solution NMR & - & 2014 \\ \hline
2N1O & 19 & Solution NMR & - & 2015 \\ \hline
2N8T & 10 & Solution NMR & - & 2016 \\ \hline
2RLY & 8 & Solution NMR & - & 2007 \\ \hline
2RM0 & 8 & Solution NMR & - & 2007 \\ \hline
2HO2 & 1 & X-ray diffraction & 1.33 & 2007 \\ \hline
2OEI & 1 & X-ray diffraction & 1.35 & 2007 \\ \hline
\end{tabular}
}
\end{table}

\begin{table}[H]
\centering
\caption{Summary of the cross-fertilization procedure pairings for the MDM2 (4HFZ) and PDZ (1ZUB) receptors. Each phase involved 10 independent docking runs for both the proper (cognate) and improper (non-cognate) peptide candidates.}
\label{tableS8}
\setlength{\tabcolsep}{8pt}
{\small
\renewcommand{\arraystretch}{1.25}
\begin{tabular}{|c|c|c|c|}
\hline
\textbf{Phase} & \textbf{Receptor (PDB ID)} & \textbf{Proper Peptide (Cognate)} & \textbf{Improper Peptide (Non-cognate)} \\ \hline
1 & MDM2 (4HFZ) & p53 segment (ETFSDLWKLLP) & ELKS1b segment (CDQDEEEGIWA) \\ \hline
2 & PDZ (1ZUB) & ELKS1b segment (CDQDEEEGIWA) & p53 segment (ETFSDLWKLLP) \\ \hline
\end{tabular}
}
\end{table}

% Mark end of Supporting Tables on the correct page
\phantomsection
\label{supp:tables:end}

\newpage

%%%%%%%%%%%%%%%%%%%%%%%%%%%%%%%%%%%%%%%%%%%%%
% ---------- Supporting Calculations ----------

\section*{Supporting Calculations}
\phantomsection
\label{supp:calcs:start}

\subsection{Phase-level ensemble means and proper-improper differences}

For each receptor, two phases are considered: a proper phase $\mathcal{P}$ and an improper phase $\mathcal{I}$.
Within a given phase $X \in \{\mathcal{P},\mathcal{I}\}$, $n_X$ independent ensembles are indexed by
$e = 1,\dots,n_X$.
Each ensemble $e$ yields: (i) an unnormalized Shannon term $S_{\Psi,e}^{'}$ (bits) computed from the macrostate probability distribution; and (ii) contact statistics that determine a phase-level normalization factor. For each phase $X$, the phase-pooled normalization is
\[
\left(\frac{Q}{M}\right)^{(X)} = \frac{Q^{(X)}}{M^{(X)}},
\]
where $Q^{(X)}$ and $M^{(X)}$ are obtained by pooling the nonzero contact pairs and their contact weights over all replicas within the phase. The reported phase-level PSI entropy is then
\[
S_{\Psi}^{\mathrm{phase},(X)} = \left(\frac{Q}{M}\right)^{(X)}\bar{S}_{\Psi}^{',(X)},
\qquad
\bar{S}_{\Psi}^{',(X)} = \frac{1}{n_X}\sum_{e \in X} S_{\Psi,e}^{'}.
\]
Phase differences are reported either as absolute differences,
\[
\Delta S_{\Psi}^{\mathrm{phase}} = S_{\Psi}^{\mathrm{phase},(\mathcal{P})} - S_{\Psi}^{\mathrm{phase},(\mathcal{I})},
\quad
\Delta\!\left(\frac{Q}{M}\right) = \left(\frac{Q}{M}\right)^{(\mathcal{P})} - \left(\frac{Q}{M}\right)^{(\mathcal{I})},
\]
or as percent differences
\[
\Delta\%\!\left(S_{\Psi}^{\mathrm{phase}}\right) =
\biggl(1 - \frac{S_{\Psi}^{\mathrm{phase},(\mathcal{P})}}{S_{\Psi}^{\mathrm{phase},(\mathcal{I})}}\biggr)\times 100\%,
\quad
\Delta\%\!\left(\frac{Q}{M}\right) =
\biggl(1 - \frac{(Q/M)^{(\mathcal{P})}}{(Q/M)^{(\mathcal{I})}}\biggr)\times 100\%.
\]

%%%%%%%%%%%%%%%%

\subsection{Standard errors for \texorpdfstring{$S_{\Psi}^{\mathrm{phase}}$}{S_Psi phase} and \texorpdfstring{$Q/M$}{Q over M}}

Ensembles are treated within a phase as independent replicas, so error bars represent the standard error (SE) of the phase mean, estimated from the across-ensemble variability.
For $S_{\Psi}^{'}$, the sample variance in phase $X$ is
\[
\sigma_{S_{\Psi}^{'}}^{2,(X)} =
\frac{1}{n_X - 1}\sum_{e \in X} \bigl(S_{\Psi,e}^{'} - \bar{S}_{\Psi}^{',(X)}\bigr)^{2},
\]
and the corresponding SE of the mean is
\[
\mathrm{SE}\!\bigl(\bar{S}_{\Psi}^{',(X)}\bigr) =
\frac{\sigma_{S_{\Psi}^{'}}^{(X)}}{\sqrt{n_X}}.
\]

For the ensemble unit-measure ratio $Q/M$, the sample variance is
\[
\sigma_{Q/M}^{2,(X)} =
\frac{1}{n_X - 1}\sum_{e \in X}
\biggl(\left(\frac{Q}{M}\right)_e - \overline{\left(\frac{Q}{M}\right)}^{(X)}\biggr)^{2},
\]
with corresponding standard error
\[
\mathrm{SE}\!\biggl(\overline{\left(\frac{Q}{M}\right)}^{(X)}\biggr)
=
\frac{\sigma_{Q/M}^{(X)}}{\sqrt{n_X}}.
\]

Error in the reported values and figures correspond to these phase-specific SE calculations for
$\bar{S}_{\Psi}^{',(X)}$ and $\overline{(Q/M)}^{(X)}$ in the proper and improper phases.

%%%%%%%%%%%%%%%%

\newpage

\subsection{Effective number of macrostates and normalized effective state fraction (phase-pooled)}

Pooling all microstates across replicas within a phase yields a phase-pooled macrostate probability distribution
\[
p_i = \frac{g_i}{\Omega}, \qquad \sum_i p_i = 1,
\]
where \(g_i\) is the pooled macrostate multiplicity and \(\Omega\) is the total number of pooled microstates. The unnormalized Shannon entropy (bits) is
\[
S_{\Psi}^{'} = -\sum_i p_i \log_2 p_i,
\]
and the corresponding Shannon effective number of macrostates (for $\log_{x}$ with base 2) is defined as
\[
N_{\mathrm{eff}} = 2^{S_{\Psi}^{'}}.
\]
For convenience, the normalized effective state fraction is reported as \(N_{\mathrm{eff}}/\Omega\), which measures the effective support of the pooled macrostate distribution relative to the total microstate count.

For the two receptors, the phase-pooled values are:
\[
\begin{array}{l|cc|cc}
 & \Omega_{\mathcal{P}} & N_{\mathrm{eff}}^{(\mathcal{P})} & \Omega_{\mathcal{I}} & N_{\mathrm{eff}}^{(\mathcal{I})} \\
\hline
4\mathrm{HFZ} & 476 & 220.5 & 501 & 258.2 \\
1\mathrm{ZUB} & 631 & 303.3 & 247 & 155.3 \\
\end{array}
\]
with corresponding normalized effective state fractions \(N_{\mathrm{eff}}/\Omega\) of 0.463 (4HFZ proper), 0.515 (4HFZ improper), 0.481 (1ZUB proper), and 0.629 (1ZUB improper).

\phantomsection
\label{supp:calcs:end}

\end{document}